\documentclass[aps,twocolumn,amsmath,amssymb,superscriptaddress,letterpaper]{revtex4}

\usepackage{times}
\usepackage{amsfonts}
\usepackage{mathrsfs}
\usepackage{graphicx}
\usepackage{dcolumn}
\usepackage{bm}
\usepackage{color,xcolor}
\usepackage[colorlinks,bookmarks=false,citecolor=blue,linkcolor=red,urlcolor=blue]{hyperref}
\usepackage{ulem}
\usepackage{multirow}
\usepackage{makecell}
\usepackage{soul}
\usepackage{diagbox}
\usepackage{physics}
\usepackage{braket}
\usepackage{array}
\usepackage{float}

\def\be{\begin{equation}}  \def\ee{\end{equation}}
\def\bea{\begin{eqnarray}}  \def\eea{\end{eqnarray}}
\def\balig{\begin{align}}   \def\ealig{\end{align}}

\def\dprime{\prime\prime}

\newcommand{\mb}[1]{\mathbf{#1}}

\begin{document}

\title{Classification of High-Ordered Topological Nodes towards Moir\'e Flat Bands in Twisted Bilayers}

\author{Fan Cui}
\affiliation{Beijing National Laboratory for Condensed Matter Physics and Institute of Physics,
Chinese Academy of Sciences, Beijing 100190, China}
\affiliation{School of Physical Sciences, University of Chinese Academy of Sciences, Beijing 100190, China}

\author{Congcong Le}
\affiliation{RIKEN Interdisciplinary Theoretical and Mathematical Sciences (iTHEMS), Wako, Saitama 351-0198, Japan}

\author{Qiang Zhang}
\affiliation{Beijing National Laboratory for Condensed Matter Physics and Institute of Physics,
Chinese Academy of Sciences, Beijing 100190, China}

\author{Xianxin Wu}
\affiliation{CAS Key Laboratory of Theoretical Physics, Institute of Theoretical Physics, Chinese Academy of Sciences, Beijing 100190, China}

\author{Jiangping Hu}\email{jphu@iphy.ac.cn}
\affiliation{Beijing National Laboratory for Condensed Matter Physics and Institute of Physics, 
Chinese Academy of Sciences, Beijing 100190, China}
\affiliation{Kavli Institute of Theoretical Sciences, University of Chinese Academy of Sciences, 
Beijing, 100190, China}
\affiliation{New Cornerstone Science Laboratory, Beijing, 100190, China}

\author{Ching-Kai Chiu}\email{ching-kai.chiu@riken.jp}
\affiliation{RIKEN Interdisciplinary Theoretical and Mathematical Sciences (iTHEMS), Wako, Saitama 351-0198, Japan}

\date{\today}

\begin{abstract}

At magic twisted angles, Dirac cones in twisted bilayer graphene (TBG) can evolve into flat bands, serving as a critical playground for the study of strongly correlated physics. When chiral symmetry is introduced, rigorous mathematical proof confirms that the flat bands are locked at zero energy in the entire Moir\'{e} Brillouin zone (BZ). Yet, TBG is not the sole platform that exhibits this absolute band flatness. Central to this flatness phenomenon are topological nodes and their specific locations in the BZ. In this study, considering twisted bilayer systems that preserve chiral symmetry, we classify various ordered topological nodes in base layers and all possible node locations across different BZs. Specifically, we constrain the node locations to rotational centers, such as $\Gamma$ and $\text{M}$ points, to ensure the interlayer coupling retains equal strength in all directions. Using this classification as a foundation, we systematically identify the conditions under which Moir\'e flat bands emerge. Additionally, through the extension of holomorphic functions, we provide proof that flat bands are locked at zero energy, shedding light on the origin of the band flatness. Remarkably, beyond Dirac cones, numerous twisted bilayer nodal platforms can host flat bands with a degeneracy number of more than two, such as four-fold, six-fold, and eight-fold. This multiplicity of degeneracy in flat bands might unveil more complex and enriched correlation physics.

\end{abstract}

\maketitle

\section{Introduction}

Twisted bilayer graphene (TBG) is a well-recognized system  to manifest Moir\'{e} flat bands (MFBs)~\cite{Bistritzer-pnas11,cao-nat18-insulator,cao-nat18}, where electron-electron correlations are strongly enhanced. Consequently, TBG  has become a compelling platform to explore diverse strong correlation phenomena, such as  Mott insulating states~\cite{Bistritzer-pnas11,cao-nat18-insulator,cao-nat18}, unconventional high-temperature superconductivity~\cite{Lu-nat19-sc_orbital_ci,christos2022,park2021tunable,cao-nat18,Yankowitz-sci19-tuning,stepanov2020}, and  quantum anomalous Hall effects \cite{Serlin-sci19-fm,Sharpe-sci19-fm,KTLaw-nc20-omagnetoelectric,tseng2022anomalous}.

The origin of the MFBs is closely related to the Dirac cones in graphene. Graphene hosts two Dirac cones at the corners of the Brillouin zone (BZ). In the twisted bilayer platform, the Dirac cones evolve to flat bands through the interlayer coupling at magic angles, exhibiting an intriguing characteristic of two-fold degeneracy per valley per spin. This connection has been further demonstrated in other similar systems with Dirac cones, such as twisted double bilayer graphene~\cite{lee_theory_2019,liu_tunable_2020,cao_tunable_2020,shen_correlated_2020}, twisted trilayer graphene~\cite{PhysRevB.87.125414,MA202118,carr_ultraheavy_2020,chen_electrically_2021,trilayer-Nature}, monolayer graphene with specific periodic potential~\cite{Advanced-Materials}, and twisted few-layer graphite~\cite{PhysRevB.100.085109,ma_moire_2022}, in which the similar type of two-fold degenerate MFBs from the Dirac cones can also be found. In contrast, literature evidence suggests that MFBs cannot originate from the twisted surfaces of 3D topological insulators possessing Dirac nodes~\cite{Stern2023,Jennifer-Cano}. Extending beyond the Dirac cone, a quadratic node at the M point of the square lattice can evolve into two-fold degenerate MFBs within the twisted bilayer framework~\cite{Yao-Hong}.
This paves the way to seek alternative twisted bilayer platforms exhibiting MFBs. On the other hand, the introduction of a spatially alternating magnetic field in TBG can lead to four-fold degenerate MFBs per valley per spin \cite{le2022double}, thereby leading to the possibility of realizing higher degeneracy of MFBs. 

This manuscript aims to identify new systems that host MFBs, explore their degeneracy greater than two, and confirm the state of absolute flatness throughout the entire Moir\'{e} BZ. In TBG, due to the preservation of chiral symmetry, it has been proved by holomorphic function extension~\cite{Ashvin-prl19-flat} that the two-fold degenerate flat bands are absolutely flat at zero energy across the entire Moir\'{e} BZ. Inspired by the band flatness deriving from Dirac and quadratic nodes, we naturally extrapolate topological nodes to $n-$ordered topological nodes protected by chiral symmetry at zero energy, such as quadratic and cubic nodes. This approach is warranted as chiral symmetry prohibits the interlayer coupling from destroying the two topological nodes. An additional requisite for TBG band flatness is the location of the Dirac cones at a $C_3$ rotation center, which ensures the interlayer momentum hopping between the two Dirac cones respects the $C_3$ rotation symmetry. In this context, we propose the extension of this rotation center condition to $C_n$, where $n\ge 3$; here, the topological nodes are positioned at the $C_n$ rotation center, and the interlayer coupling retains identical strength under the rotation. Consequently, we can classify four possible locales for an n-ordered node --- 1.~$\Gamma$ in a square lattice 2.~$M$ in a square lattice 3.~$\Gamma$ in a hexagonal lattice 4.~$\text{K},\ \text{K}^{\prime}$ in a hexagonal lattice. Using these topological $n$-ordered semimetals as a base layer, we pair two such identical layers with a minor twist to evaluate the band flatness within low-energy physics.
    
\begin{figure}[htbp]
    \centering
    \includegraphics[scale=0.35]{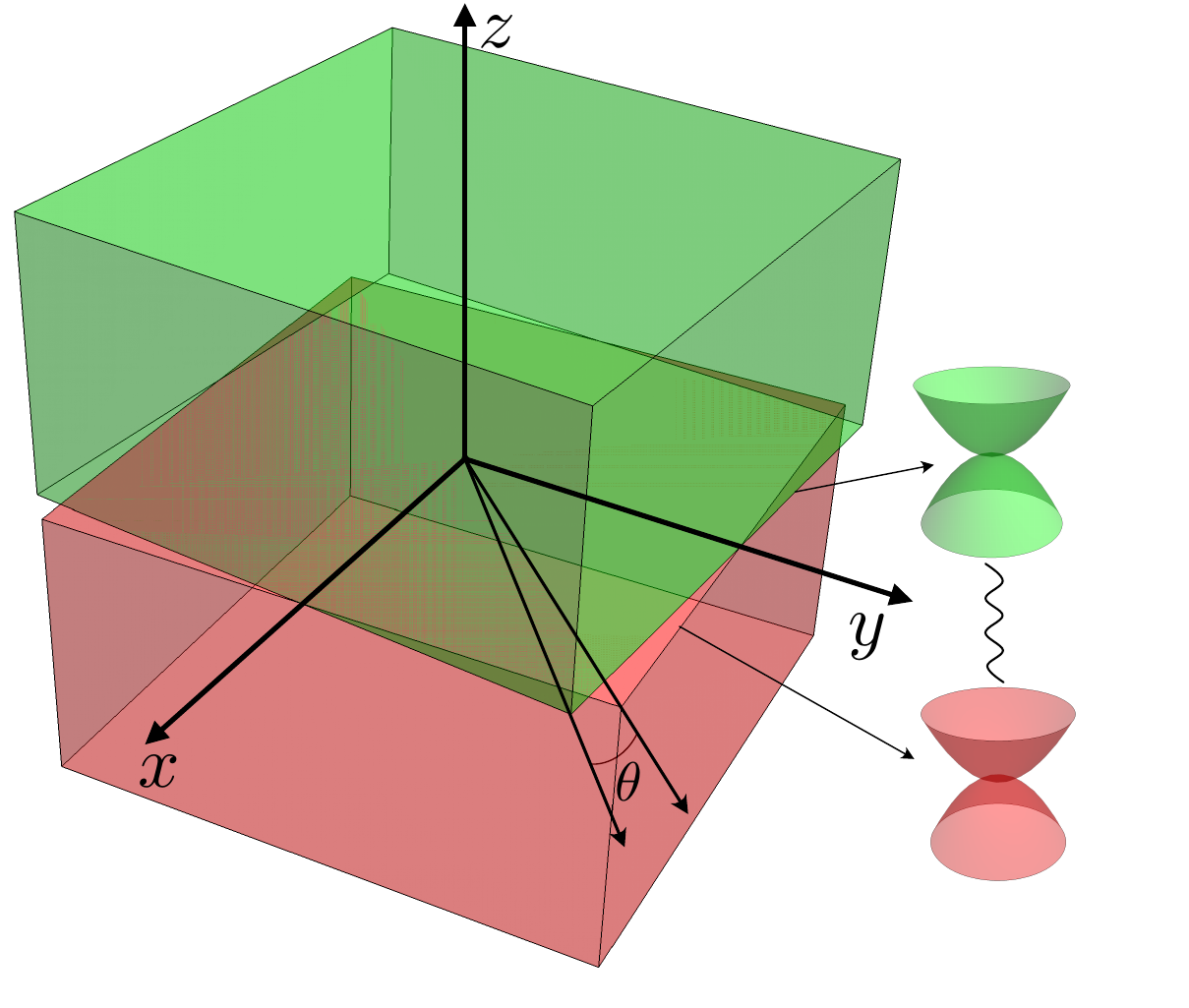}
    \caption{Two topological superconductors in contact with each other
    with a twist form a twisted bilayer platform in the interface of the surfaces. Due to the non-zero 3D winding number $W_{\rm{3D}}$, a $n$-th ordered topological node emerges on the surface. }
    \label{fig1}
\end{figure}

This manuscript primarily delves into the intricacies of twisted bilayer physics for $n-$ordered topological nodes situated at $\Gamma$ in both square and hexagonal lattices. Furthermore, we impose chiral symmetry to protect the stability of the topological nodes and lock MFBs at zero energy. These proposed twisted bilayer platforms can be realized in non-trivial 3D time-reversal symmetric topological superconductors in class DIII and CI~\cite{PhysRevB.78.195125}. The combination of particle-hole symmetry and time-reversal symmetry directly leads to chiral symmetry. With chiral symmetry, the 3D winding number $W_{\rm{3D}}$ as a topological invariant correspond to the number (order) of the topological nodes on the surface of the superconductor. An $n$-ordered node at $\Gamma/\text{M}$ on the surface BZ is one of the possible cases. Class DIII limits $n$ to be odd, while class CI limits $n$ to be even. To realize the twisted bilayer, we consider two topological superconductors with $W_{\rm{3D}}=n,\ -n$, and their two surfaces in contact with each other with a twist, as shown in Fig. \ref{fig1}. Since these two topological superconductors possess the opposite winding numbers, the two topological nodes are robust against the interlayer coupling.

The remainder of this paper is organized as follows. In Sec. \ref{section2}, we offer a concise overview of the criteria for band flatness, omitting detailed derivations for the brevity.
In Sec. \ref{section3}, we develop a generalized Hamiltonian for twisted bilayer platforms hosting generic topological nodes. We also introduce the inversion of the Birman-Schwinger operator, whose spectrum determines the magic angles. Subsequently, we provide explicit Hamiltonian expressions for both square and hexagonal lattices when a topological node is located at the $\Gamma$ point.
In Sec. \ref{section4}, we calculate the spectrum of the inverted Birman-Schwinger operator to identify these magic angles, and illustrate the energy spectrum at these angles to showcase the emergence of MFBs.
In Sec. \ref{section5}, we categorize TBSs based on the order and locations of their topological nodes, listing those that do and do not host MFBs, as well as their corresponding degeneracy numbers. In Sec. \ref{section6}, we construct the wavefunctions of these flat bands using holomorphic functions, thereby proving that these bands remain flat throughout the entire Moir\'{e} BZ. Lastly, we conclude with a summary in Sec. \ref{section7}. Some technical details have been relegated to supplementary materials.

\section{Flatness conditions at first glance}
\label{section2}

\begin{table}[tpb]
\renewcommand\arraystretch{1.5}
\setlength{\tabcolsep}{3mm}{
    \begin{tabular}{|*{2}{c|}*{5}{w{c}{1.2em}|}}
        \hline
        \multicolumn{2}{|c|}{\diagbox[width=11em]{Position}{Order}} & n=1 & n=2 & n=3 & n=4 & n=5 \\
        \hline
        \multirow{2}{*}{\makecell{Square\\Lattice}} & $\mathbf{\Gamma}$ & N & Y & Y & N & N \\
        \cline{2-7}
        & $\mathbf{M}$ & N & Y & Y & N & N \\
        \hline
        \multirow{2}{*}{\makecell{Hexagonal\\Lattice}} & $\mathbf{\Gamma}$ & N & Y & Y & Y & Y \\
        \cline{2-7}
            & $\mathbf{K}/\mathbf{K}^{\prime}$ & Y & Y & N & N & N \\
        \hline
    \end{tabular}}
    \caption{The emergence of flat bands is determined by the conditions of the topological nodes, including the order from linear to quintc and the node locations. Label 'Y/N' indicates the presence/absence of flat bands.}
    \label{Degeneracy}
\end{table}

Before delving into the technical derivations, we provide an overview of the main findings identifying the classified conditions for band flatness. As indicated by a 'Y' in Table I, specific orders $n$ and locations of topological nodes can give rise to MFBs. While the form of interlayer coupling can influence twisted bilayer physics, we simplify our model by imposing chiral symmetry and assuming that the strength of the interlayer coupling decays exponentially with distance. Moreover, we assume that the directional dependence of the interlayer coupling directly inherits characteristics from the $n$-th ordered node. These assumptions guarantee the conditions of the band flatness is independent of the detailed form of the interlayer coupling.

To contextualize our findings, we compare Table \ref{Degeneracy} with existing literature and highlight our novel contributions. Firstly, we confirm that Dirac nodes located at only $\text{K},\ \text{K}^{\prime}$ can result in flat bands, consistent with the guiding principles established for Dirac fermions~\cite{Stern2023}. Secondly, we corroborate previous work~\cite{Yao-Hong} showing that a twisted bilayer with a quadratic node at the M point can also produce flat bands. Beyond these scenarios, all cases marked with a 'Y' in Table \ref{Degeneracy} represent new results. Specifically, in the square lattice, quadratic and cubic nodes at either $\Gamma$ or $\text{M}$ can induce flat bands. In the hexagonal lattice, nodes with orders ranging from quadratic to quintic at the $\Gamma$ point can also lead to flat bands, while at $\text{K},\ \text{K}^{\prime}$, only Dirac and quadratic nodes yield flat bands. In the remainder of the manuscript, we begin by constructing a Hamiltonian for TBSs with $n$-th ordered nodes and progressively show the emergence of Moir\'{e} band flatness.

\section{general method}
\label{section3}

\begin{figure}
\centerline{\includegraphics[width=0.45\textwidth]{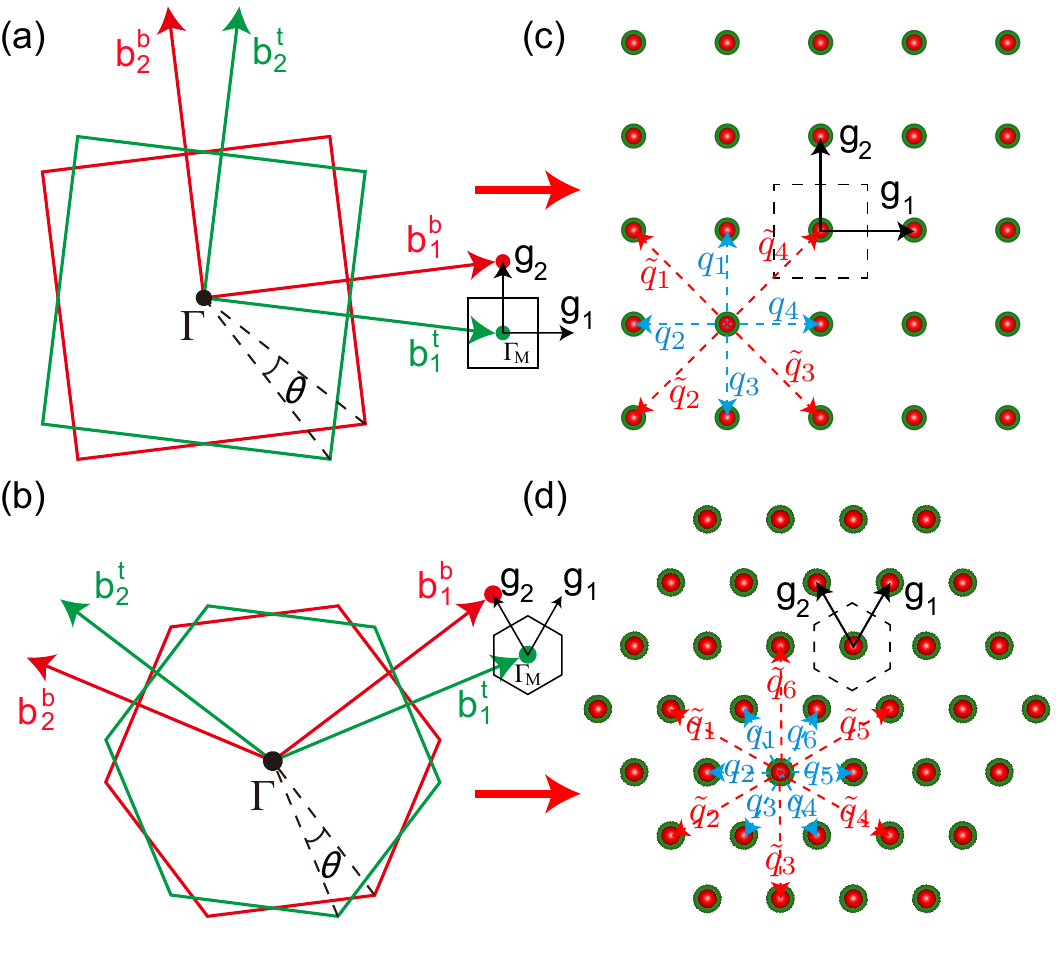}}
\caption{(color online) (a) and (b) The monolayer square and hexagonal BZs with the twist form Moir\'{e} BZs.
The red and green hexagonals represent the BZs of the bottom and top layers, each rotated by an angle $\pm\theta/2$, and the small black square and hexagonal indicates the Moir\'{e} BZs. (c) and (d) The corresponding square and hexagonal reciprocal lattice structures. The red
and green atoms indicate Dirac cones of the top and bottom layers.}
\label{fig2}
\end{figure}

\subsection{General Hamiltonian for TBSs}

We investigate twisted bilayer systems (TBSs) composed of two monolayers with a slightly twisted angle on square and hexagonal lattices. Consider the monolayers possess nodes of $n$-order at $C_n$ rotation centers $\mathbf{H}$ ($n>2$), including $\Gamma,\ \text{M}$ in the square lattice and $\Gamma,\text{K},\text{K}^{\prime}$ in the hexagonal lattice. 
The low-energy Hamiltonians near $\mathbf{H}$ can be expressed as:
\begin{equation}
\begin{aligned}
    h(\mathbf{k})
    =& v_0
    \begin{pmatrix}
    0 & (k_x-ik_y)^n \\
    (k_x+ik_y)^n & 0
    \end{pmatrix},
\end{aligned}
\label{qc_model}
\end{equation}
where $\mathbf{k}$ is a small quantity expanded from the rotation center $\mathbf{H}$, $v_0$ is a scaling parameter, and $n=1,2,3,\cdots$ for linear, quadratic, cubic or higher order nodes respectively. Different from TBG, the index of this $2\times 2$ node Hamiltonian indicates internal degree freedom, such as particle-hole, spin, orbitals or mixture \cite{PhysRevLett.102.187001}. On the contrary, the Dirac Hamiltonian in graphene is described by spatial dependent A and B bases.

The general Hamiltonian for the TBS can be constructed as
\begin{equation}
    \hat{H}=\hat{H}^t+\hat{H}^b+\hat{H}^{\perp}.
\end{equation}
Here, $\hat{H}^{t/b}$ represents the monolayer Hamiltonian for the top/bottom layer (the rotation of the monolayer Hamiltonian at a small twist angle is neglected for convenience), and $\hat{H}^{\perp}$ is the interlayer coupling term. The monolayer Hamiltonian with the $n$-ordered node can be written as:
\begin{equation}
    \hat{H}^{l}=\sum_{\mathbf{k}^{l},\alpha,\beta} \hat{\psi}^{\dagger}_{\mathbf{k}^{l},\alpha} h^{\alpha\beta}(\mathbf{k}^{l}) \hat{\psi}_{\mathbf{k}^{l},\beta},
\end{equation}
where the subscript $l=t/b$ denotes the top/bottom layer, $\alpha/\beta$ indicates the internal index. The interlayer coupling term is given by:
\begin{equation}
    \hat{H}^{\perp} = \sum^{\alpha,\beta}_{\mathbf{k}^{t},\mathbf{k}^{b}} \hat{\psi}^{\dagger}_{\mathbf{k}^{t},\alpha} T_{\mathbf{k}^t, \mathbf{k}^b}^{\alpha,\beta} \hat{\psi}_{\mathbf{k}^{b},\beta}+ h.c. \ .
    \label{intehop1}
\end{equation}
The Bloch states for the different layers are defined as:
\begin{equation}
    \ket{\mathbf{k}^l,\alpha}=\frac{1}{\sqrt{N}}\sum_{\mathbf{R}^l}e^{i(\mathbf{H}^l+\mathbf{k}^l)\cdot\mathbf{R}^l}\ket{\mathbf{R}^l,\alpha}.
    \label{Bloch state}
\end{equation}
Here $N$ denotes the total number of lattice sites, $\mathbf{H}^l$ indicates the rotation center located in layer $l$, and $\mathbf{R}^l$ is the monolayer lattice vectors. Then the matrix element $T_{\mathbf{k}^t, \mathbf{k}^b}^{\alpha,\beta} = \langle\mathbf{k}^t, \alpha|\hat{H}| \mathbf{k}^b, \beta\rangle$ in Eq. \ref{intehop1} can be written as:
\begin{equation}
    T_{\mathbf{k}^t, \mathbf{k}^b}^{\alpha,\beta} = \sum_{\mathbf{b}^t,\mathbf{b}^b} \frac{\tilde{t}_{\alpha\beta}(\mathbf{H}^t+\mathbf{k}^t+\mathbf{b}^t)}{A_{\text {u.c.}}}\delta_{\mathbf{H}^t+\mathbf{k}^t+\mathbf{b}^t,\mathbf{H}^b+\mathbf{k}^b+\mathbf{b}^b},
    \label{intehop2}
\end{equation}
which stems from the Fourier transformation of the real space interlayer hopping $\langle\mathbf{R}^t, \alpha|\hat{H}| \mathbf{R}^b, \beta\rangle$ (see the detailed derivation in the supplementary materials Sec.I). The explicit expression, which depends on the form of the node, will be discussed in Sec. \ref{subsectionC} and Sec. \ref{subsectionD}. Here, $\mathbf{b}^l$ indicates the $l$ monolayer reciprocal lattice vector, $A_{\text{u.c.}}$ indicates the unit cell area, and $\tilde{t}_{\alpha\beta}(\mathbf{H}^t+\mathbf{k}^t+\mathbf{b}^t)$ indicates the Fourier transform coefficient of the real-space interlayer hopping.

The $\delta$ function in Eq. \ref{intehop2} imposes the constraint of the hopping between the two momentum states
\begin{equation}
    \mathbf{k}^t=\mathbf{k}^b+(\mathbf{H}^b+\mathbf{b}^b)-(\mathbf{H}^t+\mathbf{b}^t).
    \label{delta-function}
\end{equation}
Therefore, this momentum hopping constraint depends on the location of the rotation center $\mathbf{H}^l$, where the topological node is located. In the main text, we focus on models with the node of different order located at $\Gamma$ in the square and hexagonal lattices, then we have $\mathbf{H}^l=\mathbf{\Gamma}^l=\mathbf{0}$, which gives
\begin{equation}
    \mathbf{k}^t=\mathbf{k}^b+\mathbf{b}^b-\mathbf{b}^t.
    \label{delta-function-for-Gamma}
\end{equation}
We leave the discussions of $\text{M}$ point in the square lattice and $\text{K},\ \text{K}^{\prime}$ points in the hexagonal lattice in the supplementary materials Sec.II. As we focus on the low energy physics at a small twited angle, we take an approximation by considering the hopping distance $|\mathbf{k}^t-\mathbf{k}^b|$ less than the length of the primitive reciprocal lattice vector. That is, the reciprocal lattice vectors in the two layers are close $\mathbf{b}^t \sim \mathbf{b}^b$ so that we can define $\delta\mathbf{b} \equiv \mathbf{b}^b - \mathbf{b}^t$. Although all of the reciprocal lattice vectors should be included in the summation, we include only $\mathbf{b}^t, \mathbf{b}^b=$ primitive reciprocal lattice vectors as the nearest neighbor hopping and $\mathbf{b}^t, \mathbf{b}^b=$ combination of the two primitive reciprocal lattice vectors as the next nearest neighbor hopping. We note that as $\mathbf{b}^t, \mathbf{b}^b=0$, the strength of the on-site coupling vanishes due to the chiral symmetry (see the explanation in the next subsection). 

Since $\mathbf{k}^l$ is small near the node, we can approximately take $\tilde{t}_{\alpha\beta}(\mathbf{H}^t+\mathbf{k}^t+\mathbf{b}^t) \approx \tilde{t}_{\alpha\beta}(\mathbf{H}^t+\mathbf{b}^t)$. The Fourier coefficient in Eq. \ref{intehop2} for the models with a node at $\Gamma$ point becomes:
\begin{equation}
    \tilde{t}_{\alpha\beta}(\mathbf{H}^t+\mathbf{k}^t+\mathbf{b}^t) 
    = \tilde{t}_{\alpha\beta}(\mathbf{k}^t+\mathbf{b}^t)
    \approx \tilde{t}_{\alpha\beta}(\mathbf{b}),
\end{equation}
where the vector $\mathbf{b}$ without a superscript denotes the untwisted reciprocal lattice vectors, we take the approximation because of the small twist angle. Then the interlayer hopping matrix can be written in an economic form
\begin{equation}
    T^{\alpha,\beta}_{\mathbf{k}^t,\mathbf{k}^b}
    =T^{\alpha\beta}_{\mathbf{k}^b+\delta\mathbf{b},\mathbf{k}^b}
    =\frac{\tilde{t}(\mathbf{b})}{A_{u.c.}}
    \equiv T^{\alpha\beta}(\delta\mathbf{b}).
    \label{intehop3}
\end{equation}
The second equality shows that under the approximation above, the interlayer hopping matrix is independent of $\mathbf{k}^l$, and only depends on the choice of $\mathbf{b}$, which gives $\delta\mathbf{b}$ correspondingly.

The repeated hopping forms a momentum space lattice, as shown in Fig. \ref{fig2} (c) and (d) and the difference $\delta\mathbf{b}$ between the primitive reciprocal lattice vectors in the two layers forms a much smaller primitive reciprocal lattice vector of the Moir\'{e} BZ, which are denoted by $\mathbf{g}_i$, as shown in Fig. \ref{fig2} (a) and (b). The size of the Moir\'{e} BZ can be characterized by the magnitude of the vectors $\mathbf{g}_i$, which is written as
\begin{equation}
    |\mathbf{g}_i|=2|\mathbf{b}_i|\sin{\frac{\theta}{2}} ,
\end{equation}
where $|\mathbf{b}_i|$ is the magnitude of the monolayer primitive reciprocal lattice vectors.
For momentum $\mathbf{k}^l$ outside the Moir\'{e} BZ, the momentum can be expressed as $\mathbf{k}^l=\mathbf{k}+\mathbf{g}$, where $\mathbf{k}$ is defined in the Moir\'{e} BZ, and $\mathbf{g}=m \mathbf{g}_1+n \mathbf{g}_2$ is a Moir\'{e} reciprocal lattice vector, as illustrated in Fig. \ref{fig2} (c) and (d). Then the general Hamiltonian for the twisted bilayer system can be rewritten in the form of the BZ folding
\begin{equation}
\begin{aligned}
    \hat{H} =& \sum^{{l},\alpha,\beta}_{\mathbf{k},\mathbf{g}} \hat{\psi}^{l\dagger}_{\mathbf{k}+\mathbf{g},\alpha} h^{\alpha\beta}(\mathbf{k}+\mathbf{g}) \hat{\psi}^l_{\mathbf{k}+\mathbf{g},\beta}
    \\
    &+ \sum^{\alpha,\beta}_{\mathbf{k},\mathbf{g},\delta\mathbf{b}} \hat{\psi}^{t\dagger}_{\mathbf{k}+\mathbf{g}+\delta\mathbf{b},\alpha} T^{\alpha\beta}(\delta\mathbf{b}) \hat{\psi}^b_{\mathbf{k}+\mathbf{g},\beta}+ h.c. \ .
\end{aligned}
\label{Hamiltonian}
\end{equation}
Furthermore, the general Hamiltonian $\hat{H}$ can be economically written as $\hat{H}=\sum_{\mathbf{k}}\hat{H}(\mathbf{k})$, where $\hat{H}(\mathbf{k})$ reads
\begin{equation}
\begin{aligned}
    \hat{H}(\mathbf{k})&=\hat{H}^t(\mathbf{k})+\hat{H}^b(\mathbf{k})+\hat{H}^{\perp}(\mathbf{k}) ,
    \\
    \hat{H}^l(\mathbf{k})&=\sum^{\alpha,\beta}_{\mathbf{g}} \hat{\psi}^{l\dagger}_{\mathbf{k}+\mathbf{g},\alpha} h^{\alpha\beta}(\mathbf{k}+\mathbf{g}) \hat{\psi}^l_{\mathbf{k}+\mathbf{g},\beta} ,
    \\
    \hat{H}^{\perp}(\mathbf{k})&=\sum^{\alpha,\beta}_{\mathbf{g},\delta\mathbf{b}} \hat{\psi}^{t\dagger}_{\mathbf{k}+\mathbf{g}+\delta\mathbf{b},\alpha} T^{\alpha\beta}(\delta\mathbf{b}) \hat{\psi}^b_{\mathbf{k}+\mathbf{g},\beta}+ h.c. \ ,
\end{aligned}
\label{total Hamiltonian}
\end{equation}
where $H^l(\mathbf{k})$ represents the monolayer Hamiltonian or the on-site terms in the momentum space lattice, while $H_{\perp}(\mathbf{k})$ represents the hopping term between momentum space lattice sites with the hopping matrix $T(\delta\mathbf{b})$. As the nodal points of the top and bottom layers are located at the $\Gamma$ point, they have the same sites in the momentum space, denoted by the red and green dots as shown in Fig. \ref{fig2} (c,d).

Since the unit cell is enlarged to the Moir\'e super unit cell and $\mathbf{k}$ shrinks to the Moir\'{e} BZ, we use $\mathbf{r}$ to describe the real space position vector in a super unit cell. 
The annihilation operator in this basis can be written as
\begin{equation}
    \hat{\psi}^{l}_{\mathbf{k},\alpha}(\mathbf{r})=\frac{1}{\sqrt{N}}\sum_{\mathbf{g}}e^{i\mathbf{g}\cdot\mathbf{r}}\hat{\psi}^{l}_{\mathbf{k}+\mathbf{g},\alpha} ,
\end{equation}
where $N$ is the total number of momentum sites. In this regard, the Hamiltonian can be expressed in terms of $\mathbf{r}$
\begin{equation}
\begin{aligned}
    \hat{H}_{\mathbf{k}}^l({\mathbf{r}})&=\sum_{\alpha,\beta} \hat{\psi}^{l\dagger}_{\mathbf{k},\alpha}(\mathbf{r}) h^{\alpha \beta}\left(\mathbf{k}-i\mathbf{\nabla}_{\mathbf{r}}\right) \hat{\psi}^{l}_{\mathbf{k},\beta}(\mathbf{r}) ,
    \\
    \hat{H}_{\mathbf{k}}^{\perp}({\mathbf{r}})&= \sum_{\alpha,\beta} \hat{\psi}^{t\dagger}_{\mathbf{k},\alpha}(\mathbf{r}) T^{\alpha\beta}(\mathbf{r}) \hat{\psi}^{b}_{\mathbf{k},\beta}(\mathbf{r})+ h.c. \ ,
\end{aligned}
\end{equation}
where the matrix forms of $h(\mathbf{k}-i\mathbf{\nabla}_{\mathbf{r}})$ and $T(\mathbf{r})$ are given by 
\begin{equation}
\begin{aligned}
    & h(\mathbf{k}-i\mathbf{\nabla}_{\mathbf{r}})
    =
    \begin{pmatrix}
    0 & (k-i2{\partial})^n \\
    (\bar{k}-i2\bar{\partial})^n & 0
    \end{pmatrix} ,
    \\
    & T(\mathbf{r})=\sum_{\delta\mathbf{b}} e^{-i\delta\mathbf{b} \cdot \mathbf{r}}
    \begin{pmatrix}
        0 & T_{12}(\delta\mathbf{b}) \\
        T_{21}(\delta\mathbf{b}) & 0
    \end{pmatrix}.
\end{aligned}
\end{equation}
To simplify the notations, we define $\partial\equiv \frac{1}{2}(\partial_x-i\partial_y), \bar{\partial}\equiv\frac{1}{2}\left(\partial_x+i \partial_y\right)$ and $k\equiv\mathbf{k}_x-i\mathbf{k}_y, \bar{k}\equiv\mathbf{k}_{x}+i\mathbf{k}_{y}$. Due to chiral symmetry, the diagonal terms in $T(\mathbf{r})$ must vanish. Hence, in the real space, the chirally symmetric Hamiltonian $H_{\mathbf{k}}(\mathbf{r})$ in the basis of $\Phi_{\mathbf{k}}(\mathbf{r})=\left(\phi^{t}_{\mathbf{k}}(\mathbf{r}), \chi^{t}_{\mathbf{k}}(\mathbf{r}), \phi^{b}_{\mathbf{k}}(\mathbf{r}), \chi^{b}_{\mathbf{k}}(\mathbf{r})\right)^{\top}$ reads:
\begin{equation}
    H_{\mathbf{k}}(\mathbf{r}) =
    \begin{pmatrix}
    h(\mathbf{k}-i\nabla_{\mathbf{r}}) & T(\mathbf{r}) \\
    T^{\dagger}(\mathbf{r}) & h(\mathbf{k}-i\nabla_{\mathbf{r}})
    \end{pmatrix} .
\end{equation}
Then, by reshuffling the basis to $\Phi_{\mathbf{k}}(\mathbf{r})=\left(\phi^{t}_{\mathbf{k}}(\mathbf{r}), \phi^{b}_{\mathbf{k}}(\mathbf{r}), \chi^{t}_{\mathbf{k}}(\mathbf{r}), \chi^{b}_{\mathbf{k}}(\mathbf{r})\right)^{\top}$, the chirally symmetric Hamiltonian becomes
\begin{equation}
\begin{aligned}
    H_{\mathbf{k}}(\mathbf{r})&=
    \begin{pmatrix}
        0 & D_{\mathbf{k}}^{*}(-\mathbf{r})   \\
        D_{\mathbf{k}}(\mathbf{r}) &  0
    \end{pmatrix} ,
    \\
    D_{\mathbf{k}}(\mathbf{r})&=
    \begin{pmatrix}
        (\bar{k}-i2\bar{\partial})^n
        &
        \sum_{\delta{\mathbf{b}}}e^{-i\delta{\mathbf{b}} \cdot \mathbf{r}}T_{21}(\delta\mathbf{b})
        \\
        \sum_{\delta{\mathbf{b}}}e^{i\delta{\mathbf{b}} \cdot \mathbf{r}}T^*_{12}(\delta\mathbf{b})
        &
        (\bar{k}-i2\bar{\partial})^n
    \end{pmatrix} .
\label{real model}
\end{aligned}
\end{equation}
We obtain the chirally symmetric general Hamiltonian for TBSs with the node located at the $\Gamma$ point. Once the interlayer hopping matrix is settled for specific models, the energy spectrum can be calculated. Now we proceed to discuss the interlayer hopping terms.

\subsection{Momentum hopping matrix $T(\delta\mathbf{b})$}

Finding the explicit matrix expression of the interlayer momentum hopping $T(\delta\mathbf{b})$ is the key to compute the low energy physics of the TBSs. The momentum hopping matrix $T(\delta\mathbf{b})$ can be obtained via Fourier transformation of the interlayer hopping in real space, as expressed by the following equation:
\begin{equation}
    \tilde{t}_{\alpha\beta}(\mathbf{p}) = \int e^{-i \mathbf{p} \cdot \mathbf{r}} t_{\alpha\beta}(\mathbf{r}) d \mathbf{r}.
    \label{Fourier coefficient 1}
\end{equation}
Here, we assume the real space interlayer hopping $t_{\alpha\beta}(\mathbf{r})$ depends on the direction and inherits the rotation symmetry from the $n$-order nodes, which can be written as
\begin{equation}
    t_{\alpha\beta}(\mathbf{r})=t(r)(U^\dagger_{\theta_{\mathbf{r}}}{\mathbf \sigma} U_{\theta_{\mathbf{r}}})_{\alpha\beta}
    \label{realhop},
\end{equation}
where $U_{\theta_{\mathbf{r}}}=e^{i\theta_{\mathbf{r}}J_z\sigma_z}$  denotes the rotation symmetry involving angular momentum $J_z\equiv n/2$, and ($r$, $\theta_{\mathbf{r}}$) representing the polar coordinates of vector $\mathbf{r}$. We assume that the strength of the interlayer coupling exhibits an exponential decay $t(r)=e^{-{r}/{r_0}}$, with $r_0$ set to 1 for convenience. (We will discuss later that different decay rates may affect the values of the magic angles but cannot make magic angles vanish or appear. That is, the form of the decay does not affect the emergence of MFBs.). In general, ${\mathbf \sigma}=a\sigma_x+b\sigma_y+c\sigma_z+d\sigma_0$ represents the hopping matrix in the $\theta_{\mathbf{r}}=0$ direction. Using the expression of Eq. \ref{realhop}, the hopping matrix in momentum space can be written in the form of
\begin{equation}
    \tilde{t}(\mathbf{p})=
    \begin{pmatrix}
        (d+c)D(\mathbf{p}) & (a+bi)e^{-2i J_z\theta_{\mathbf{p}}}C_{J_z}(\mathbf{p}) \\
        (a-bi)e^{2i J_z\theta_{\mathbf{p}}}C_{J_z}(\mathbf{p}) & (d-c)D(\mathbf{p})
    \end{pmatrix} ,
    \label{realhop2}
\end{equation}
where 
\begin{equation}
\begin{aligned}
    & D(\mathbf{p})=\int e^{-ipr\cos\theta} t(r)rdrd\theta , \\
    & C_{J_z}(\mathbf{p})=\int e^{-ipr\cos\theta} t(r) e^{\pm 2i J_z\theta} rdrd\theta ,
\end{aligned}
\label{integral}
\end{equation}
with $r$/$\theta$ ranging from 0/0 to $+\infty$/$2\pi$.

\begin{figure}
\centerline{\includegraphics[width=0.5\textwidth]{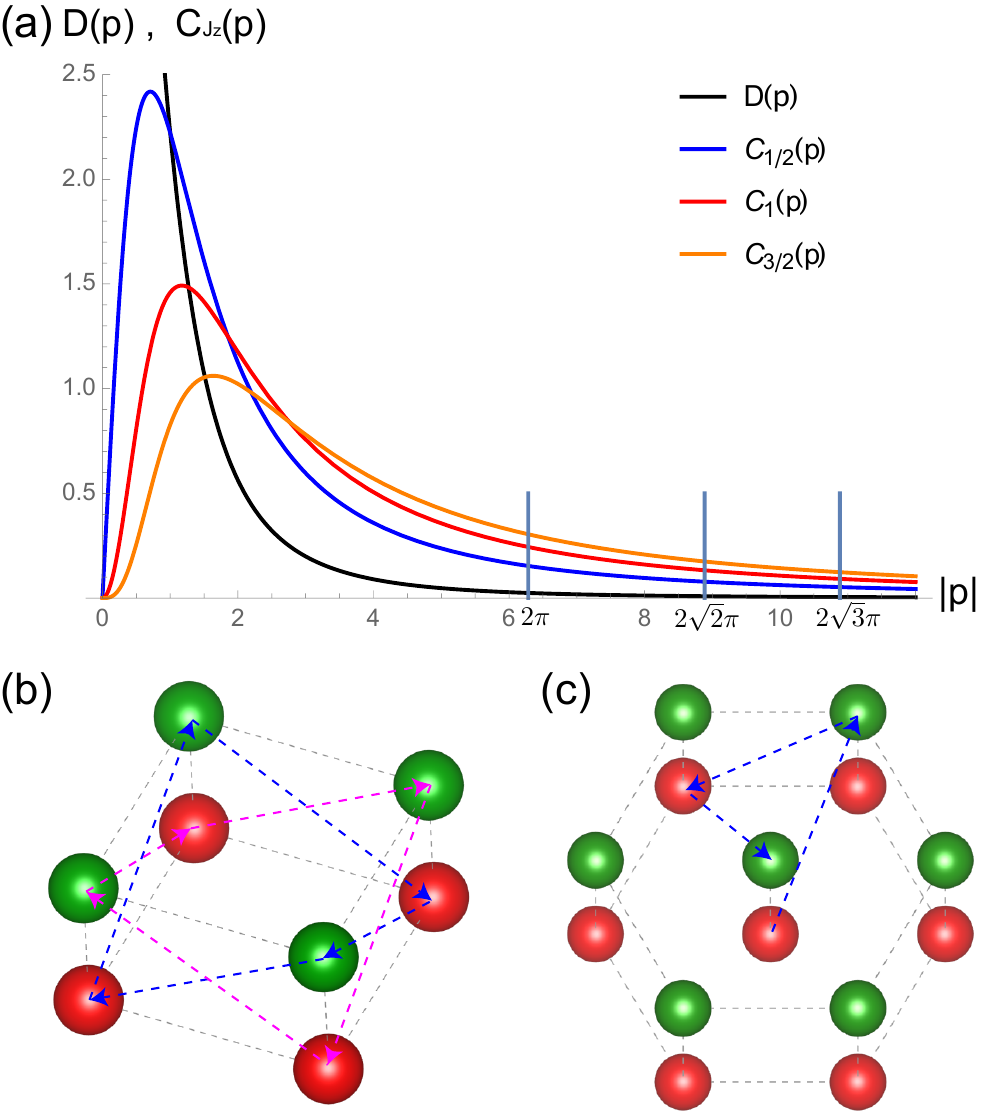}}
\caption{(a) The functions $D(\mathbf{p})$ and $C_{J_z}(\mathbf{p})$ on $|\mathbf{p}|$ for linear ($J_z=\frac{1}{2}$), quadratic ($J_z=1$) and cubic node ($J_z=\frac{3}{2}$) describing the strength of the interlayer coupling. (b) In the presence of the NN momentum hopping (dashed arrows), two decoupled momentum lattices in the square lattice. (c) Momentum lattice connected by NN hopping in the hexagonal lattice.\label{fig3} }
\end{figure}

Fig. \ref{fig5}(a) illustrates the function $D(\mathbf{p})$ labeled by black line. Because chiral symmetry is preserved, we neglect this diagonal term. At $\theta_{\mathbf{r}}=0$, the phase of the off-diagonal term aligns with the $n$-ordered node Eq. \ref{qc_model}. Since $C_{J_z}(\mathbf{p})$ is real, $b$ must vanish. Then the interlayer hopping matrix in momentum space becomes
\begin{equation}
\begin{aligned}
    \tilde{t}(\mathbf{p}) = & aC_{J_z}(\mathbf{p})
    \begin{pmatrix}
            0 & e^{-2i J_z\theta_{\mathbf{p}}} \\
        e^{2i J_z\theta_{\mathbf{p}}} & 0 
    \end{pmatrix}.
    \label{realhop3}
\end{aligned}
\end{equation}
Additionally, Fig. \ref{fig3}(a) also shows $C_{J_z}(\mathbf{p})$ on momentum $|\mathbf{p}|$ for linear node ($J_z=\frac{1}{2}$), quadratic node ($J_z=1$) and cubic node ($J_z=\frac{3}{2}$) denoted by blue, red and orange lines, respectively. These lines exhibit zero values at $|\mathbf{p}|=0$, indicating that the on-site momentum hopping matrix $T(0)$ in Hamiltonian Eq. \ref{total Hamiltonian} vanishes due to chiral symmetry. Furthermore, $C_{J_z}(\mathbf{p})$ grow rapidly then decay slowly, with relatively large values around $|\mathbf{p}|=|\mathbf{b}_1|$ and $\sqrt{2}|\mathbf{b}_1| / \sqrt{3}\mathbf{b}_1|$ (for the square/hexagonal lattice respectively). Hence, besides nearest-neighbor (NN), we also need to consider the next nearest-neighbor (NNN) momentum hoppings in the square and hexagonal lattices for a good approximation.

\subsection{Chirally symmetric Hamiltonian in the square lattice}
\label{subsectionC}

\begin{figure}
    \centering
    \includegraphics[scale=0.35]{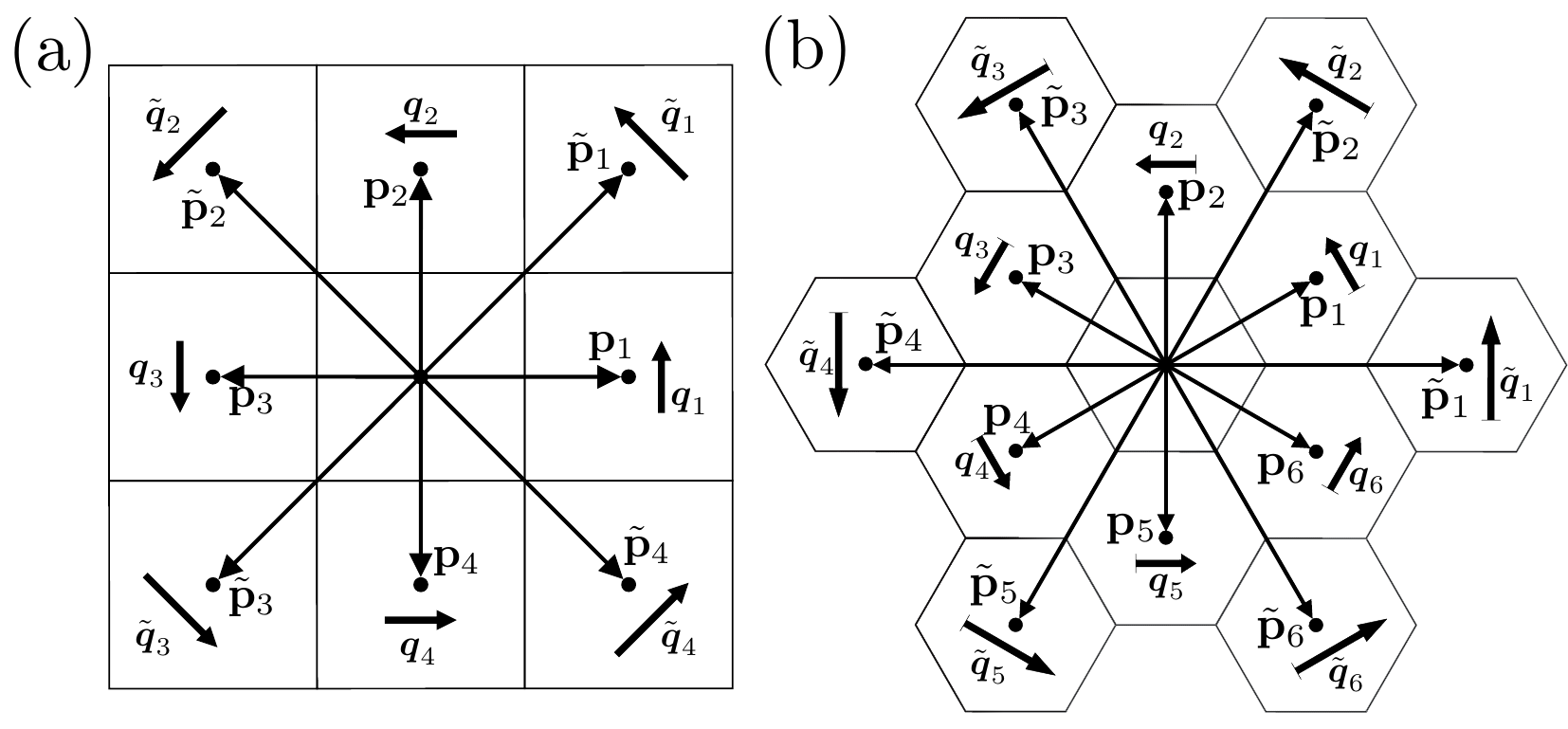}
    \caption{The reciprocal vectors $\mathbf{b}=\mathbf{p}_i, \tilde{\mathbf{p}}_j$ and the corresponding $\delta\mathbf{b}=\mathbf{q}_i, \tilde{\mathbf{q}}_j$ for the square and hexagonal lattices. The angles of the vectors $\theta_{\mathbf{p}_i}$ and $\theta_{\tilde{\mathbf{q}}_j}$ can be easily seen from this figure.}
    \label{fig4}
\end{figure}

Knowing the strength of the interlayer coupling as a function of momentum, we can derive an explicit expression for the twisted bilayer Hamiltonian. We focus on dominant momentum hoppings to construct the interlayer coupling, while constraining the momenta by \(\mathbf{k}^t = \mathbf{k}^b + \mathbf{b}^b - \mathbf{b}^t\). In the square lattice, four primitive reciprocal lattice vectors $\mathbf{b}^l=\pm\mathbf{b}_1^{l}$, $\pm\mathbf{b}_2^{l}$ respectively form four NN momentum hoppings, while the four combinations of the primitive reciprocal lattice vectors $\mathbf{b}^l=\mathbf{b}_1^{l}\pm\mathbf{b}_2^{l}$, $-\mathbf{b}_1^{l}\pm\mathbf{b}_2^{l}$ form four NNN momentum hoppings. Due to the small twisted angle, we disregard layer label $l$ and mark these two types of the four vectors by $\mathbf{p}_i$ and $\tilde{\mathbf{p}}_j$ respectively. Furthermore, the corresponding momentum differences $\delta\mathbf{b}$ between the two layers denoted as $\mathbf{q}_i$ and  $\tilde{\mathbf{q}}_j$ as shown in Fig. \ref{fig2}(c) and Fig. \ref{fig4}(a). 
Since angles $\theta_{\mathbf{p}_i}=(i-1)\frac{\pi}{2}$ and $\theta_{\tilde{\mathbf{p}}_j}=\frac{\pi}{4}+(i-1)\frac{\pi}{2}$ for $\mathbf{q}_i$ and $\tilde{\mathbf{q}}_j$, the explicit expression $\tilde{t}(\mathbf{p})$ of the interlayer coupling can be written in Eq. \ref{realhop3} as $\mathbf{p}=\mathbf{p}_i,\ \tilde{\mathbf{p}}_j$. 

As we assume the lattice constant is the unity and the decay length of the interlayer coupling is also the unity, the ratio of the NN hopping and NNN hopping $\lambda_{J_z}^s\equiv C_{J_z}(|\mathbf{b}^t_1+\mathbf{b}^t_2|)/C_{J_z}(|\mathbf{b}^t_1|)$ can be fixed. In particular, $\lambda_1^s\approx 0.54$ for the interlayer coupling inheriting the direction dependence of the quadratic node. {Again, it will be shown later that changing the ratio, which only moves the values of the magic values, does not affect the emergence of the flat bands.  }

Based on real space Hamiltonian Eq. \ref{real model}, the chirally symmetric Hamiltonian $H^\eta_{\mathbf{k}}(\mathbf{r})$ can be given by
\begin{equation}
    \begin{aligned}
        H^\eta_{\mathbf{k},J_z}(\mathbf{r})&=
        &\begin{pmatrix}
            0 & D_{\mathbf{k},J_z}^{\eta*}(-\mathbf{r}) \\
            D^\eta_{\mathbf{k},J_z}(\mathbf{r}) & 0
        \end{pmatrix} ,
        \\
        D^\eta_{\mathbf{k},J_z}(\mathbf{r})& =
        &\begin{pmatrix}
            (\bar{k}-i2\bar{\partial})^n &  \alpha^\eta_{J_z} U^\eta_{J_z}(\mathbf{r}) \\
            \alpha^l_{J_z} U^\eta_{J_z}(-\mathbf{r}) & (\bar{k}-i2\bar{\partial})^n
        \end{pmatrix} ,
        \label{model_S}
    \end{aligned}
\end{equation}
where we define normalized coupling strength $\alpha^\eta_{J_z}\equiv \frac{aC(|\mathbf{b}^t_1|)}{v_0k^n_{\theta}}$ and $k_{\theta}=2|\mathbf{b}_1|\sin\frac{\theta}{2}$. Only the NN hoppings and the NNN hoppings are included to describe the interlayer coupling 
\be
U^\eta_{J_z}(\mathbf{r})=2U^{\eta\text{NN}}_{J_z}(\mathbf{r})+2U^{\eta\text{NNN}}_{J_z}(\mathbf{r}). 
\ee
We let $\eta=s$ to represent the square lattice and later $\eta=h$ is used for the hexagonal lattice. For the quadratic node ($J_z=1$), the explicit forms of the momentum hoppings are written as
\begin{equation}
\begin{aligned}
    &U^{s\text{NN}}_1(\mathbf{r})=\cos(\mathbf{q}_1\cdot\mathbf{r})-\cos(\mathbf{q}_2\cdot\mathbf{r}) , \\
    &U^{s\text{NNN}}_1(\mathbf{r})=0.54i(\cos(\tilde{\mathbf{q}}_1\cdot\mathbf{r})-\cos(\tilde{\mathbf{q}}_2\cdot\mathbf{r})) .
\end{aligned}
\end{equation}
If the NNN hopping potential $U^{s\text{NNN}}_1(\mathbf{r})$ is ignored and only the $U^{s\text{NN}}_1(\mathbf{r})$ is preserved in the Hamiltonian $H^s(\mathbf{r})$ shown in Fig. \ref{fig2}(c), the momentum lattice describing this system can be divided into two decoupled momentum space lattices. One is linked by blue dotted lines with arrows, while the other one is connected by pink dotted lines with arrows, as shown in Fig. \ref{fig3}(b). Hence, the twisted bilayer system always exhibits at least 2-fold degeneracy. When the NNN hoppings, the 2-fold degeneracy is lifted.

\subsection{Chirally symmetric Hamiltonian in the hexagonal lattice}
\label{subsectionD}

Following the similar discussions in the square lattice, we can obtain an explicit form of the Hamiltonian in the hexagonal lattice. The six primitive reciprocal lattice vectors $\pm\mathbf{b}_1^{l}$, $\pm\mathbf{b}_2^{l}$, $\pm(\mathbf{b}_1^{l}+\mathbf{b}_2^{l})$ give rise to six nearest-neighbor (NN) momentum hoppings. Meanwhile, the six combinations of these primitive vectors, $\pm(\mathbf{b}_1^{l}-\mathbf{b}_2^{l})$, $\pm(2\mathbf{b}_1^{l}+\mathbf{b}_2^{l})$, $\pm(\mathbf{b}_1^{l}+2\mathbf{b}_2^{l})$ form six next-nearest-neighbor (NNN) momentum hoppings. Likewise, we denote these two sets of six vectors as \(\mathbf{p}_i\) and \(\tilde{\mathbf{p}}_j\), respectively. Moreover, the corresponding momentum differences \(\delta \mathbf{b}\) between the two layers are denoted as \(\mathbf{q}_i\) and \(\tilde{\mathbf{q}}_j\), as illustrated in Fig. \ref{fig2}(d) and Fig. \ref{fig4}(b). For \(\mathbf{q}_i\) and \(\tilde{\mathbf{q}}_j\), the angular variables \(\theta_{\mathbf{p}_i}  = (i-1)\frac{\pi}{2}\) and \(\theta_{\tilde{\mathbf{p}}_j} = \frac{\pi}{4} + (i-1)\frac{\pi}{2}\) allow us to express the interlayer coupling \(\tilde{t}(\mathbf{p})\) explicitly, as given in Eq. \ref{realhop3} for \(\mathbf{p} = \mathbf{p}_i, \tilde{\mathbf{p}}_j\).

Fixing the lattice constant and the decay length of the interlayer coupling as the unity, we can fix the ratio  $\lambda_{J_z}^h\equiv C_{J_z}(|\mathbf{b}^t_1-\mathbf{b}^t_2|)/C_{J_z}(|\mathbf{b}^t_1|)$. Particularly, for the quadratic node we have, $\lambda_{1}^h =C_{1}(|\mathbf{b}^t_1-\mathbf{b}^t_2|)/C_{1}(|\mathbf{b}^t_1|)\approx 0.355$, for the cubic node we have $\lambda_{\frac{3}{2}}^h =C_{\frac{3}{2}}(|\mathbf{b}^t_1-\mathbf{b}^t_2|)/C_{\frac{3}{2}}(|\mathbf{b}^t_1|)\approx 0.37$. The Hamiltonian $H^h_{\mathbf{k},J_z}(\mathbf{r})$ for quadratic and cubic nodes in the hexagonal lattice has the same form as $H^\eta_{\mathbf{k},J_z}(\mathbf{r})$ in Eq. \ref{model_S}, where $\eta=h$ represents the hexagonal lattice.
Particularly, for the quadratic node, the two dominant momentum hopping can be written as
\begin{equation}
\begin{aligned}
        U^{h\text{NN}}_1(\mathbf{r}) =& (\frac{1}{2} + i\frac{\sqrt{3}}{2}) \cos(\mathbf{q}_1\cdot\mathbf{r}) - \cos(\mathbf{q}_2\cdot\mathbf{r})  \\
    &+ (\frac{1}{2}-i\frac{\sqrt{3}}{2})\cos(\mathbf{q}_3\cdot\mathbf{r}),
    \\
    U^{h\text{NNN}}_1(\mathbf{r}) =& 0.37(\cos(\tilde{\mathbf{q}}_1\cdot\mathbf{r})-(\frac{1}{2} - i\frac{\sqrt{3}}{2})\cos(\tilde{\mathbf{q}}_2\cdot\mathbf{r})  \\
    &-(\frac{1}{2}+i\frac{\sqrt{3}}{2})\cos(\tilde{\mathbf{q}}_3\cdot\mathbf{r})).
\end{aligned}
\end{equation}
Likewise, for the cubic node, we have
\begin{equation}
\begin{aligned}
    & U^{h\text{NN}}_{\frac{3}{2}}(\mathbf{r}) =-i(\sin(\mathbf{q}_1\cdot\mathbf{r})-\sin(\mathbf{q}_2\cdot\mathbf{r})+\sin(\mathbf{q}_3\cdot\mathbf{r})) ,
    \\
    & U^{h\text{NNN}}_{\frac{3}{2}}(\mathbf{r}) = -0.355(\sin(\tilde{\mathbf{q}}_1\cdot\mathbf{r})-\sin(\tilde{\mathbf{q}}_2\cdot\mathbf{r})+\sin(\tilde{\mathbf{q}}_3\cdot\mathbf{r})) .
\end{aligned}
\end{equation}
Unlike the twisted bilayer system of the square lattice, the hexagonal lattice does not exhibit 2-fold degeneracy as the NNN hoppings vanish. As the twisted bilayer Hamiltonian is established, we can introduce an approach to identify magic angles.

\subsection{The spectrum of magic angles }
\label{spectrum}

We investigate the conditions for the existence of absolutely zero-energy MFBs in the above three chirally symmetric Hamiltonians. For the convenience, we express these Hamiltonians in a unified and economic form
\begin{equation}
\begin{aligned}
    H_{\mathbf{k}}(\mathbf{r})&=&
    \begin{pmatrix}
    0 & D_{\mathbf{k}}^{*}(-\mathbf{r}) \\
    D_{\mathbf{k}}(\mathbf{r}) &  0  \\
    \end{pmatrix} ,
\end{aligned}
\label{model1}
\end{equation}
where 
\begin{equation}
\begin{aligned}
    D_{\mathbf{k}}(\mathbf{r})&=&
    \begin{pmatrix}
        (\bar{k}-i2\bar{\partial})^n &  \alpha U(\mathbf{r})     \\
        \alpha U(-\mathbf{r}) &  (\bar{k}-i2\bar{\partial})^n
    \end{pmatrix} .
\end{aligned}
\label{model_all}
\end{equation} 
Our goal is to look for the magic values of $\alpha$ leading to the MFBs, since $\alpha$ is a tunable parameter controlled by the twisted angle. We rewrite the off-diagonal term 
\begin{equation}
    D_{\mathbf{k}}(\mathbf{r})=(\bar{k}-i2\bar{\partial})^n(I-\alpha T_{\mathbf{k}}) ,\nonumber
\end{equation}
where we define
\begin{equation}
    T_{\mathbf{k}}\equiv -{(\bar{k}-i2\bar{\partial})^{-n}}
    \begin{pmatrix}
    0 & U(\mathbf{r}) \\
    U(-\mathbf{r}) & 0
    \end{pmatrix} ,
\label{DT}
\end{equation}
which is known as the Birman-Schwinger operator~\cite{Simon2021}. The emergence of absolutely zero-energy MFBs in these systems is characterized by the determinant of its Hamiltonian $H(\mathbf{r})$ being zero for any $\mathbf{k}$, i.e. $\det(D_{\mathbf{k}}(\mathbf{r}))=0$. For non-zero $\textbf{k}$, the matrix $[I-\alpha T_{\mathbf{k}}]$ must have at least one zero eigenvalue due to $\det(\bar{k}-i2\bar{\partial})^n\ne 0$, and two low-energy states at $\mathbf{k}=0$  are fixed at zero energy. According to Ref.~\cite{Simonv4,Simon2021}, for TBG the eigenvalues of $T_{\mathbf{k}}$ are independent of $\textbf{k}$. Presumably, we extend this $\textbf{k}$-independence to our generic bilayer  systems, and then define a spectrum $\mathcal{A}=1 /\operatorname{Spec}\left(T_{\mathbf{k}}\right)$ and a corresponding two-component eigenstate $\psi^0_{\mathbf{k}}(\mathbf{r})$. Consequently, by tuning the parameter $\alpha$ to one of the real eigenvalues in $\mathcal{A}$, we obtain $D_{\mathbf{k}}(\mathbf{r})\psi^0_{\mathbf{k}}(\mathbf{r})=0$ for any $\mathbf{k}$, which leads to the emergence of zero-energy MFBs. In this regard, we can compute complex spectrum $\mathcal{A}$ to determine the magic values of $\alpha$. Although by introducing alternative magnetic field $\alpha$ can gain a magnetic phase to become a complex number~\cite{le2022double}, we focus on real $\alpha$ here.

Furthermore, we demonstrate that the degeneracy of MFB is twice that of
the corresponding real eigenvalue for spectrum $\mathcal{A}$. To prove this, we investigate the eigenfunction of Birman-Schwinger operator $T_{\mathbf{k}}$ with a real eigenvalue $\eta$ independent of $\mathbf{k}$, which reads
\begin{equation}
    T_{\mathbf{k}} \phi^m_{\mathbf{k}}(\mathbf{r})=\eta \phi^m_{\mathbf{k}}(\mathbf{r}) ,
    \label{Tq_eigenvector}
\end{equation}
where $\phi^m_{\mathbf{k}}(\mathbf{r})$ are corresponding eigenvectors with m-fold degeneracy. Then, by using Eq. \ref{DT} and $\alpha=\frac{1}{\eta}$, we derive the eigenfunction of $D_{\mathbf{k}}(\mathbf{r})$ to be
\begin{equation}
    D_{\mathbf{k}}(\mathbf{r}) \phi^m_{\mathbf{k}}(\mathbf{r})=(\bar{k}-i2\bar{\partial})^n(I-\frac{1}{\eta} T_{\mathbf{k}}) \phi^m_{\mathbf{k}}(\mathbf{r})=0 ,
\end{equation}
which imply that $\phi^m_{\mathbf{k}}(\mathbf{r})$ are zero-energy eigenvectors of $D_{\mathbf{k}}(\mathbf{r})$ with $m$-fold degeneracy. Simiarly, the zero energy eigenfunction of $D^*_{\mathbf{k}}(-\mathbf{r})$ with $m$-fold degeneracy is given by $D^*_{\mathbf{k}}(-\mathbf{r}) \phi^{*m}_{\mathbf{k}}(-\mathbf{r})=0$. By exploiting the relation between $D_{\mathbf{k}}(\mathbf{r})$ and Hamiltonian $H_{\mathbf{k}}(\mathbf{r})$ in Eq. \ref{model_all}, we obtain the zero energy eigenstates of Hamiltonian $H_{\mathbf{k}}(\mathbf{r})$ with $2m$-fold degeneracy in the form of 
\begin{equation}
    \Psi^m_{\mathbf{k},1}(\mathbf{r})=
    \begin{pmatrix}
    \phi^m_{\mathbf{k}}(\mathbf{r}) \\
    0
    \end{pmatrix},
    ~~
    \Psi^m_{\mathbf{k},2}(\mathbf{r})=
    \begin{pmatrix}
    0 \\
    \phi^{*m}_{\mathbf{k}}(-\mathbf{r})
    \end{pmatrix} .
    \label{H_eigenvector}
\end{equation}
Consequently, based on Eq. \ref{Tq_eigenvector} and Eq. \ref{H_eigenvector}, we conclude that the degeneracy of the MFB is twice that of the corresponding real eigenvalue for spectrum $\mathcal{A}$.

\begin{figure}
\centerline{\includegraphics[width=0.5\textwidth]{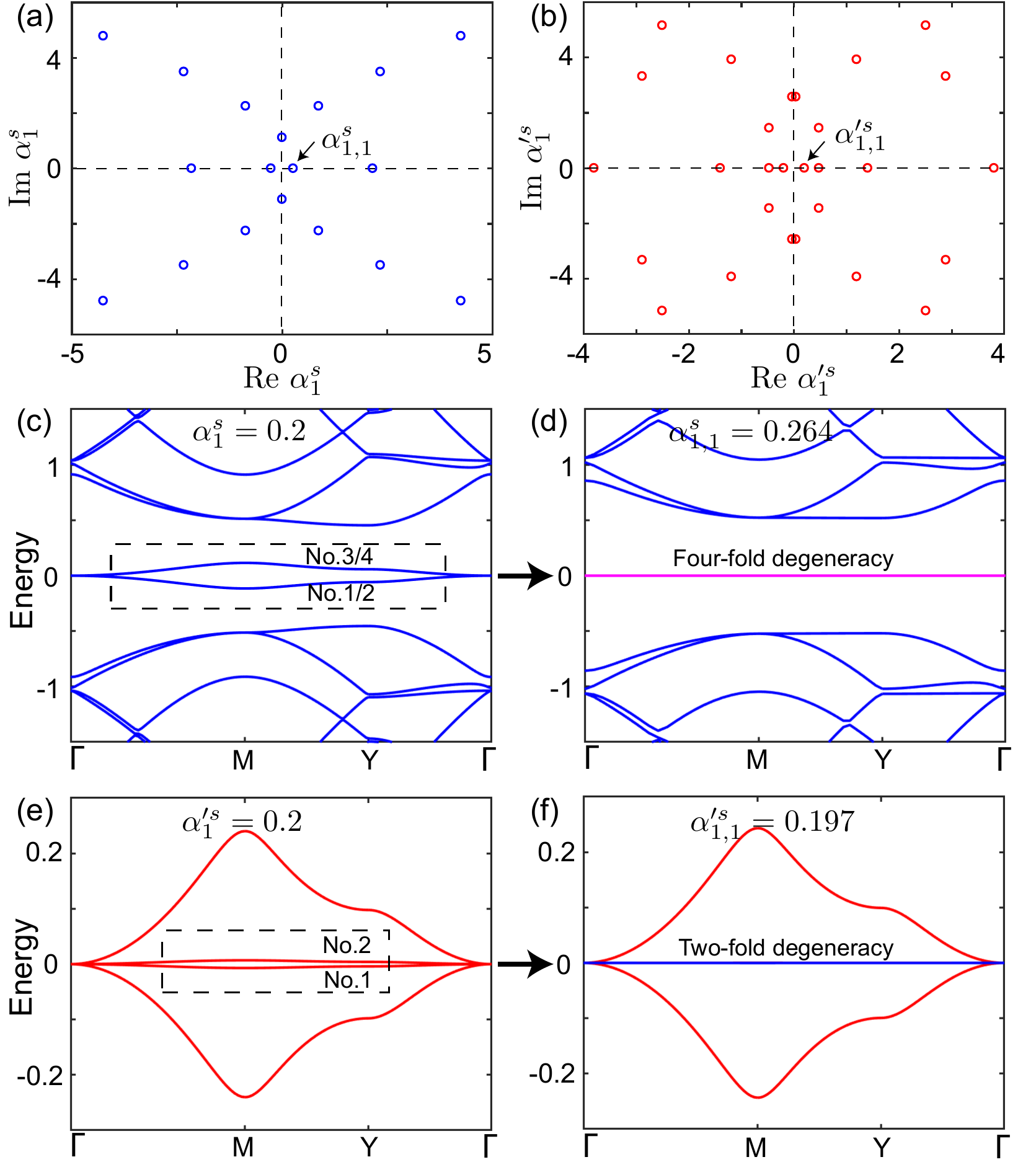}}
\caption{(color online) Spectra $\mathcal{A}$ and band structures of the Hamiltonian $H^{s}(\mathbf{r})$ in the square lattice. (a) and (b) The eigenvalues $\alpha^s_1$ and $\alpha^{\prime s}_1$ with and without the NNN hopping term. The band structures without the NNN hopping term at (c) $\alpha^s_1=0.2$; (d) First magic value $\alpha^s_{1,1}=0.264$ marked by black arrow in (a). The band structures with the NNN hopping term at (c) $\alpha^{\prime s}_1=0.2$; (d) First magic value $\alpha^{\prime s}_{1,1}=0.197$ marked by black arrow in (b).}
\label{fig5}
\end{figure}

\section{Search for flatness}
\label{section4}

To pinpoint the magic values of \(\alpha\), we initially calculate \(\mathcal{A} = \frac{1}{\text{Spec}\left(T_{\mathbf{k}}\right)}\) for a specified momentum in the Moir\'{e} BZ using Eq. \ref{DT}. The real spectrum of \(\mathcal{A}\) should correspond to \(\alpha\) values that yield MFBs, assuming that the spectrum \(\mathcal{A}\) is invariant with respect to \(\mathbf{k}\).

To validate this assumption and scrutinize the band flatness, we use the real eigenvalues of \(\mathcal{A}\) as \(\alpha\) in the twisted bilayer Hamiltonian described by Eq.~\ref{real model}. This methodology enables us to plot the energy spectrum as a function of momentum \(\mathbf{k}\) within the Moir\'{e} BZ. Employing this approach, we demonstrate that these real eigenvalues lead to MFBs at zero energy in twisted bilayers for both square and hexagonal lattices. Later, we will offer a rigorous proof that substantiates the complete flatness of these bands across the entire Moir\'{e} BZ, achieved through the extension of holomorphic functions.

\subsection{Quadratic nodes in the square lattice}

We start with a quadratic node at the $\Gamma$ point of a square lattice. For simplification, we initially neglect the second nearest momentum hopping ($U^{s\text{NNN}}_1(\mathbf{r})=0$).
This allows the Hamiltonian $H^s(\mathbf{r})$ to be cleanly divided into two identical sub-Hamiltonians, as depicted in Fig.~\ref{fig3}(b). As a result, the system inherently possesses a two-fold degeneracy. This characteristic leads the eigenvalues $\alpha^s_{1,i}$ of the spectrum $\mathcal{A}$ to exhibit a two-fold degeneracy, represented by blue circles in Fig. \ref{fig5}(a).
Since the eigenvalues continuously appear on the real axis, the MFBs emerge at extensive magic values of $\alpha^s_{1,i}$. By Eq. \ref{H_eigenvector}, the MFBs have four-fold degeneracy for each magic values, even when presented as complex numbers. 
Fig. \ref{fig5}(c) shows near the zero energy two two-fold degenerate dispersive bands, as indicated by band No.1-4 in the black dotted box; remarkably, these four bands evolve four-fold degenerate (quadruple) flat bands (magenta line) in the energy gap at the first magic value ($\alpha^s_{1,1}=0.264$) as shown in Fig. \ref{fig5}(d). 

Non-trivial topology emerges in the degenerate flat bands. As discussed in supplementary materials Sec.III, within the quadruple flat bands, the two occupied bands No.(1$\sim$2) have a total Chern number of $+2$ while the two unoccupied bands No.(3$\sim$4) possess an opposite Chern number in total. 

Now we recover the second nearest momentum hopping, which is determined by the ratio between the NNN and NN hoping strengths ($\lambda_1^s=0.54$); the two identical sub-Hamiltonians merges to one. Each two-fold degenerate state in spectrum $\mathcal{A}$ splits up into two non-degenerate states $\alpha'^s_{1,i}$ indicated red circles in Fig. \ref{fig5}(b) and the real eigenvalues still stay on the real axis. Meanwhile no additional eigenvalues appear on the real axis. The corresponding magic values lead to two-fold degenerate (double) MFBs. Similarly, the splitting from two-fold degenerate states to two non-degenerate states also occurs at energy bands. For example, the energy band degeneracy splitting is demonstrated by the transition from Fig. \ref{fig5}(c) to (e) with non-zero $U^{s\text{NNN}}_1(\mathbf{r})$. In Fig. \ref{fig5}(e), two (No.1 and 2) of the four non-degenerate dispersive bands near Fermi level evolve to double MFBs at the first magic value ($\alpha^{\prime s}_{1,1}=0.192$), as shown by the blue line in Fig. \ref{fig5} (f). Furthermore, the MFBs, which connects to other energy bands, are gapless. 

\begin{figure}
\centerline{\includegraphics[width=0.5\textwidth]{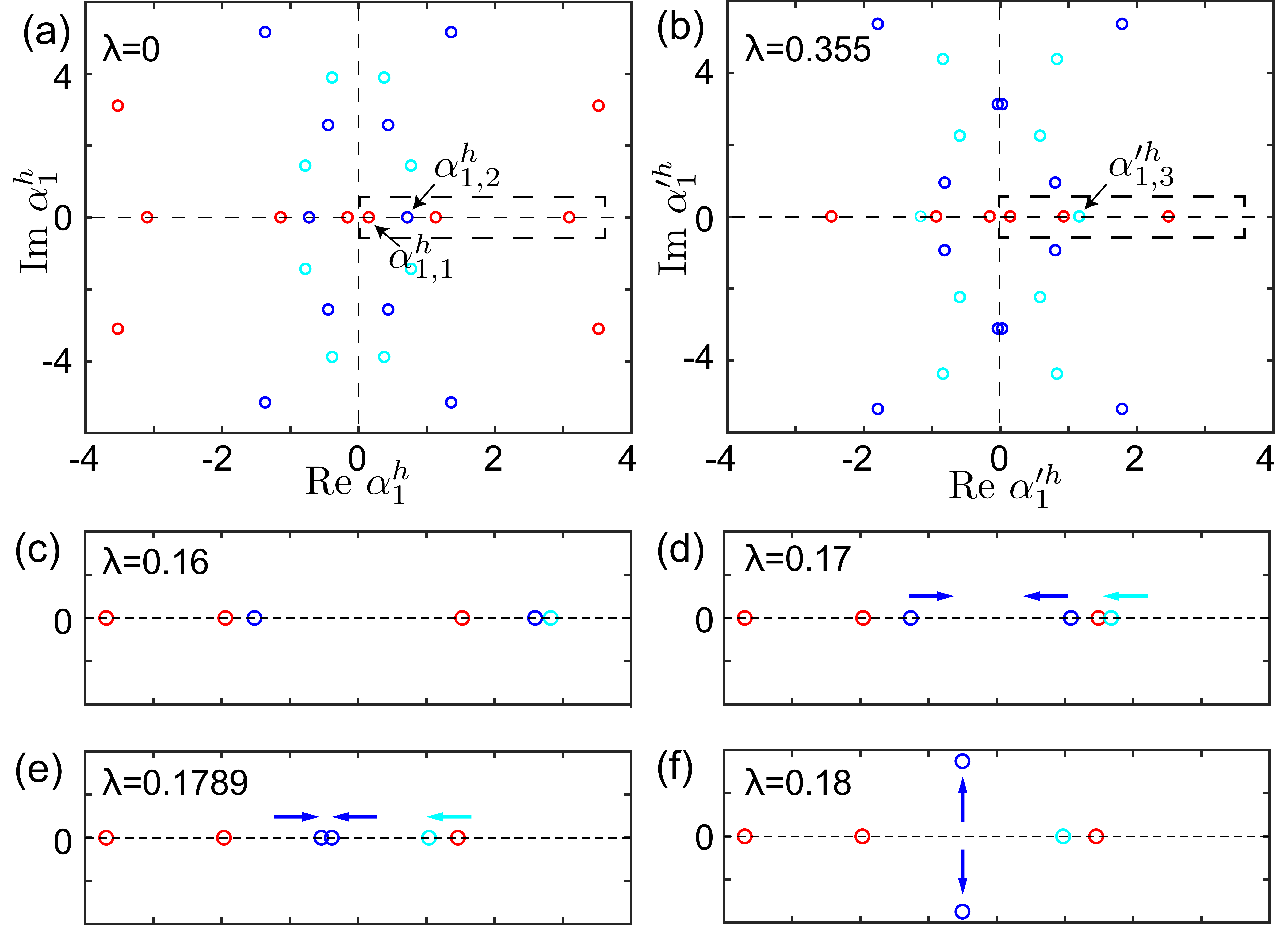}}
\caption{(color online) Spectra $\mathcal{A}$ from quadratic nodes in the hexagonal lattice. (a) and (b) The eigenvalues $\alpha^h_1$ and $\alpha^{\prime h}_1$ with and without the NNN hopping term. The evolution of magic values at different $\lambda$: (c) $\lambda=0.16$; (d) $\lambda=0.17$; (e) $\lambda=0.1789$; (f) $\lambda=0.18$. }
\label{fig6}
\end{figure}

\subsection{Quadratic and cubic nodes in the hexagonal lattice}

In this section, we examine the spectrum $\mathcal{A}$ and the emergence of highly degenerate MFBs for the chirally symmetric Hamiltonians $H^h_1(\mathbf{r})$ and $H^h_{\frac{3}{2}}(\mathbf{r})$ in the hexagonal lattice.

For the twisted bilayer of the quadratic nodes ($J_z=1$), we present eigenvalues $\alpha^h_{1,i}$ in spectrum $\mathcal{A}$ without the NNN hopping as shown in Fig. \ref{fig6}(a). Non-degenerate states (red circles) and two-fold degenerate states (blue circles) scatter particularly on the real axis; they respectively indicates two-fold and four-fold degenerate MFBs. After turning on the NNN hopping ($\lambda_1^h=0.355$), we plot eigenvalues $\alpha'^h_{1,i}$ in spectrum $\mathcal{A}$ in Fig. \ref{fig6}(b). Three-fold degenerate states (cyan circles) appear on the real axis and represents six-fold degenerate MFBs. Introducing the NNN hopping leads to the movement of spectrum $\mathcal{A}$, especially on the real axis. First, as $\lambda^h_1$ increases from zero, the eigenvalue of the triple degenerate states move from a large real number to a small number as demonstrated in Fig. \ref{fig6}(c,d). Second, with further increase to $\lambda^h_1=0.18$, the eigenvalues of the two two-fold degenerate states collide on the real axis and move away from the real axis as the complex conjugation partners as shown in Fig. \ref{fig6}(e,f). 

\begin{figure}
\centerline{\includegraphics[width=0.5\textwidth]{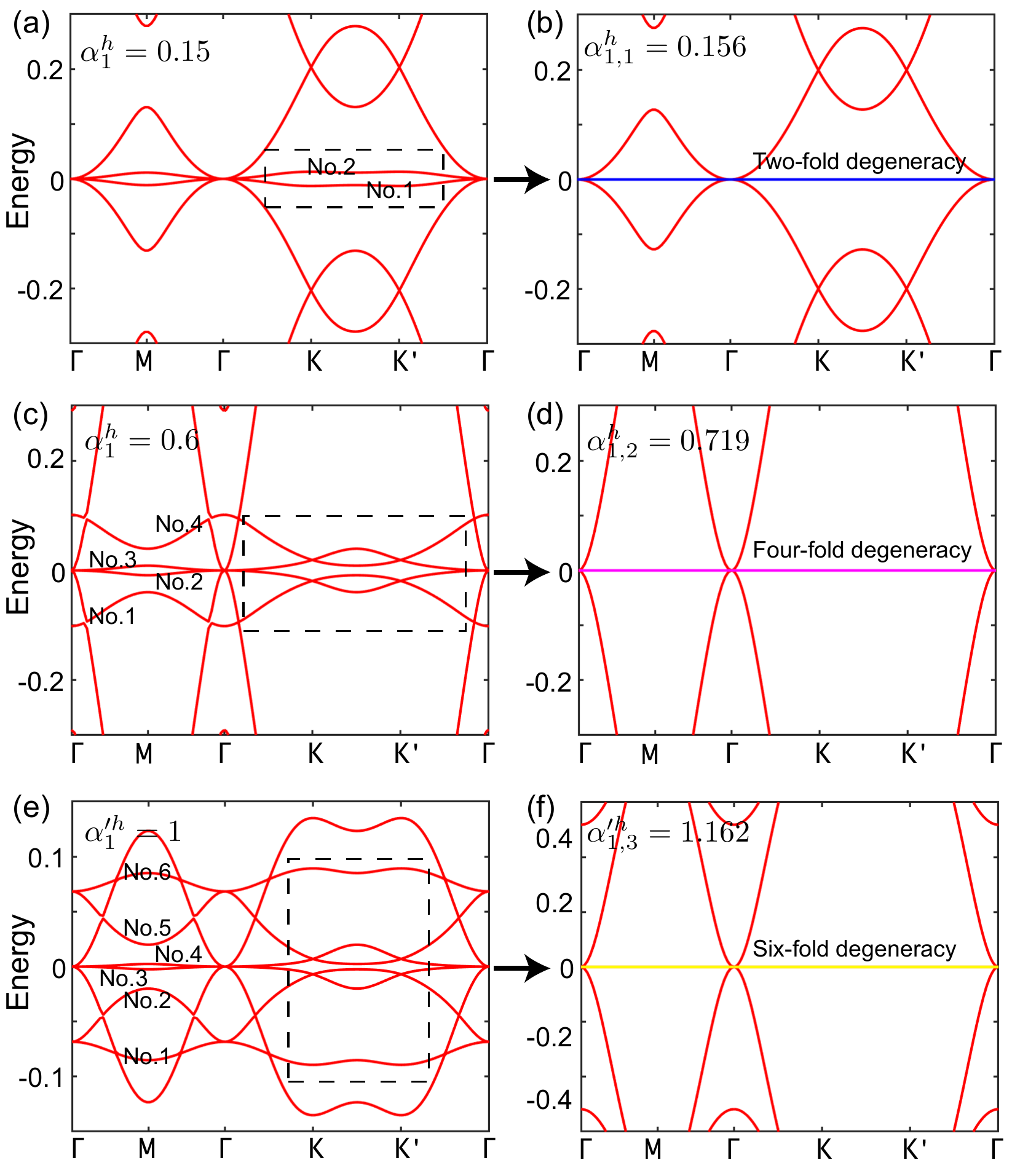}}
\caption{(color online) Band structures of Hamiltonian $H^h_1(\mathbf{r})$ at different values in the hexagonal lattice: (a) $\alpha^h_1=0.15$; (b) First magic value $\alpha^h_{1,1}=0.156$; (c) $\alpha^h_1=0.6$; (d) Second magic value $\alpha^h_{1,2}=0.719$; (e) $\alpha^{\prime h}_1=1$; (f) The third magic value $\alpha^{\prime h}_{1,3}=1.162$.
\label{fig7} }
\end{figure}

We pick up the interlayer strength ($\alpha^h_1$) near and at eigenvalues in spectrum $\mathcal{A}$ to draw energy spectra in the Moir\'{e} BZ. For a non-magic value of $\alpha^h_1=0.15$, Fig. \ref{fig7}(a) showcases four dispersive bands proximate to the Fermi level. Notably, two of these bands (No.1 and 2) flatten completely at the first magic value, $\alpha^h_{1,1}=0.156$, as indicated by the blue line in Fig. \ref{fig7}(b). Upon increasing $\alpha^h_1$ to $0.6$, Fig. \ref{fig7}(c) depicts six dispersive bands near the Fermi level, with four of these bands (No.1-4) becoming absolutely flat at the second magic value, $\alpha^h_{1,2}=0.719$, highlighted by the magenta line in Fig. \ref{fig7}(d). When $\alpha^{\prime h}_1=1$, Fig. \ref{fig7}(e) reveals eight dispersive bands close to the Fermi level. At the third magic value, $\alpha^{\prime h}_{1,3}=1.162$, six of these bands (No.1-6) are entirely flat, as demonstrated in Fig. \ref{fig7}(f).  In all three scenarios, a gapless feature is evident, as the dispersive bands connect the zero-energy flat bands at $\Gamma$.

\begin{figure}
\centerline{\includegraphics[width=0.5\textwidth]{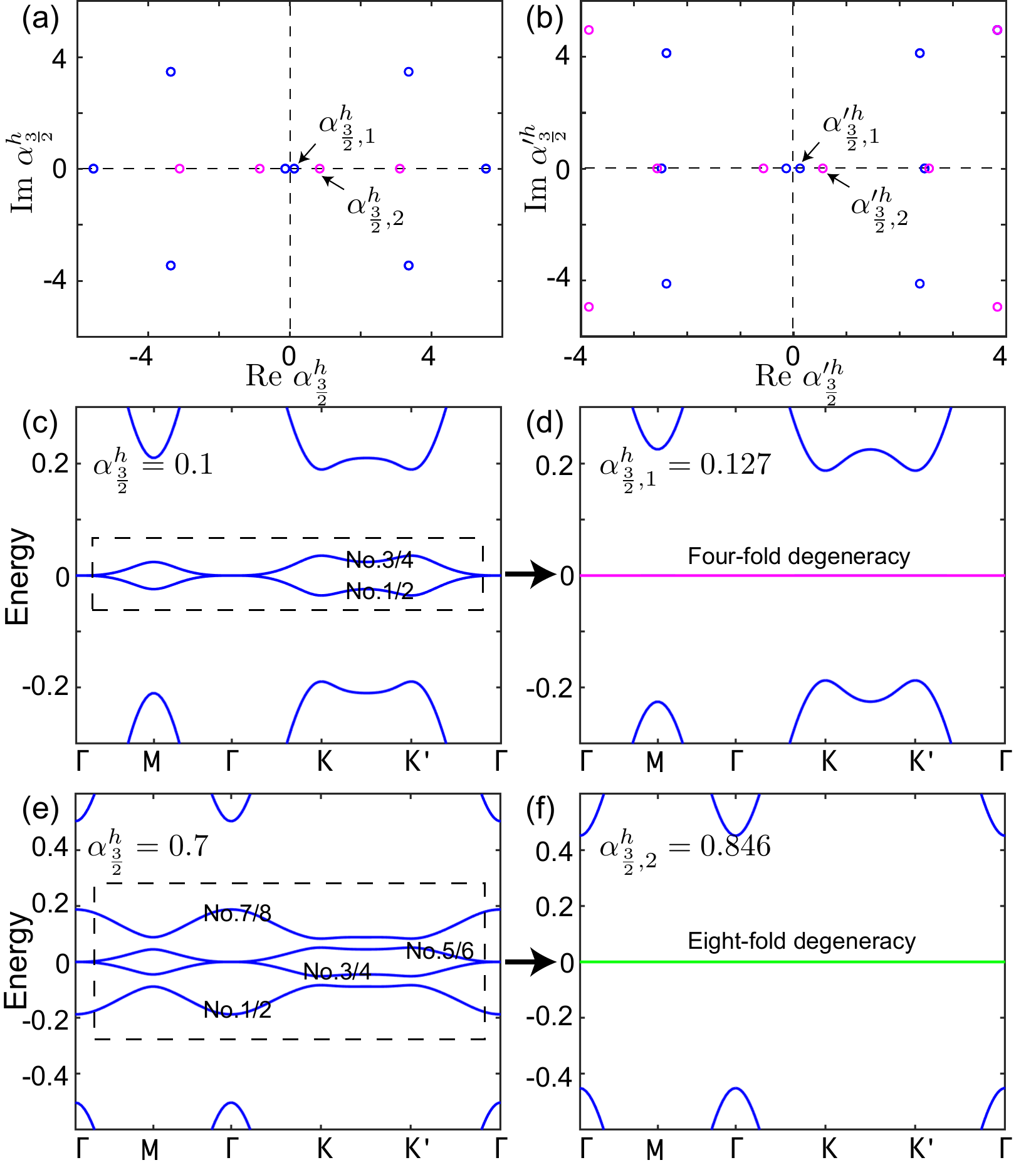}}
\caption{(color online) Spectra $\mathcal{A}$ and band structures of Hamiltonian $H^h_{\frac{3}{2}}(\mathbf{r})$ in the hexagonal lattice. (a) and (b) The eigenvalues $\alpha^h_{\frac{3}{2}}$ and $\alpha^{\prime h}_{\frac{3}{2}}$ with and without the NNN hopping term. The band structures at different value: (c) $\alpha^h_{\frac{3}{2}}=0.1$; (d) First magic value $\alpha^h_{\frac{3}{2},1}=0.127$ marked by black arrow in (a); (e) $\alpha^h_{\frac{3}{2}}=0.7$; (f) Second magic value $\alpha^h_{\frac{3}{2},1}=0.127$ marked by black arrow in (a).}
\label{fig8}
\end{figure}

Considering cubic nodes ($J_z=3/2$) in the hexagonal lattice, we respectively plot eigenvalues $\alpha^h_{\frac{3}{2},i}$ and $\alpha'^h_{\frac{3}{2},i}$ in spectrum $\mathcal{A}$ with and without the NNN hoping term. For the NNN hoping term, we choose the ratio between the NNN and the NN hopping to be $\lambda^h_{\frac{3}{2}}= 0.37$. In both cases, the eigenvalues corresponding to only double (blue circles) and quadruple (magenta circles) degenerate states can be located on the real axis; therefore, the MFBs can emerge at magic values ($\alpha^h_{\frac{3}{2},i}$, $\alpha'^h_{\frac{3}{2},i}$) with four-fold and eight-fold degeneracies. That is, the number of the MFB degeneracy is always multiple of 4. The reason is that the entire twisted bilayer system preserves an effective space-time symmetry and the operator obey $(PT)^2=-1$ (see the details in supplementary materials Sec.IV); therefore, the Kramers' theorem leads to the two-fold degenerate spectrum in the Moir\'{e} BZ in general. On the other hand, the presence of the NNN  hopping just rearrange and move from the eigenvalues $\alpha^h_{\frac{3}{2},i}$ to $\alpha'^h_{\frac{3}{2},i}$ on the real axis. In other words, considering the NN hopping can determine the existence of the MFBs. 

Let us demonstrate two examples for the evolutions of the highly degenerate MFBs. At the non-magic value $\alpha^h_{\frac{3}{2}}=0.1$, Fig. \ref{fig8}(c) shows two dispersive bands with two-fold degeneracy near the Fermi level (black dotted box). Interestingly, these four bands (No.1-4) become absolutely flat at the first magic value $\alpha^h_{\frac{3}{2},1}=0.127$, as shown in Fig. \ref{fig8}(d). By further increasing $\alpha^h_{\frac{3}{2}}=0.7$, Fig. \ref{fig8}(e) displays four dispersive bands with two-fold degeneracy near the Fermi level. At the second magic value $\alpha^h_{\frac{3}{2},2}=0.846$, all eight bands (No.1-8) become absolutely flat, as shown in Fig. \ref{fig8}(f). In both instances, the MFBs reside within energy gaps. Similarly, when accounting for NNN hopping, the gaps are still open with the MFBs inside for the first two magic values. 

The flat bands at the first two magic values inherent the non-trivial topology from the cubic node ($n=3$). As discussed in supplementary materials Sec. III, within the four-fold and eight-fold degenerate flat bands, the occupied bands No.(1$\sim$2) and No.(1$\sim$4) have the same total Chern numbers of +3, while the unoccupied bands No.(3$\sim$4) and No.(5$\sim$8) possess the opposite total Chern numbers. Moreover, a $\mathbb{Z}_{2}$ topological invariant emerges in class DIII in a 2D system. Due to the odd Chern number, the $\mathbb{Z}_{2}$ topology is non-trivial.

\section{Summary Table for Band Flatness}
\label{section5}

A fundamental insight from the spectrum calculation reveals that considering only NN hopping is sufficient to identify the emergence of the MFBs. It should be noted that for higher-ordered nodes, the momentum hopping decays at a notably slower rate as the momentum distance increases (refer to Eq. \ref{integral} and Fig. \ref{fig3}(a)).
While extending the analysis to higher-ordered nodes may necessitate the inclusion of longer-range momentum hopping for accurate approximations, it is reasonable to posit that taking into account merely the NN hopping will  be enough to determine the existence of the MFBs for different ordered nodes. Moreover, we also integrate the consideration of the NNN hopping to further confirm the MFB criteria. 

Building upon the calculation approach above, we employ the forms of the Hamiltonian (Eq.~\ref{real model}) and the interlayer hopping (Eq.~\ref{realhop3}) to catalog the properties of the MFBs for higher ordered nodes at $\Gamma/\text{M}$ point in square and $\Gamma/\text{K}/\text{K}^\prime$ in the hexagonal lattice. 
Table \ref{MFBtable} shows the MFBs emerge as $n=2, 3$ for the square lattice and as $n=1, 2, 3, 4, 5$ for the hexagonal lattice. Here we consider the first four magic values for each case, which is sufficient to show the characteristics of each model. Within the table, the numerical values signify the numbers of MFB degeneracies, while the accompanying subscripts categorize the MFBs according to whether they are gapped from or connected to other bands, denoted by $g/c$, respectively.

There are several features of the MFBs we would like to point out. First, the square lattice with only the NN hopping can be described by two identical sub-Hamiltonians, leading to two-fold degeneracy for a generic energy band. Hence, the degeneracy number of the MFBs in the Moir\'{e} BZ is always a multiple of four. After the NNN hopping is introduced, the degeneracy symmetry is broken. The MFB degeneracy is reduced to two-fold but when $n$ is odd, the degeneracy is still four-fold. Second, As $n$ is odd, space-time inversion symmetry is preserved with $(PT)^2=-1$. According to Kramer's theorem, each energy band is two-fold degenerate so that the number of the MFB degeneracy is always a multiple of four, too. Third, even without symmetries, MFBs with higher degeneracy can also arise at some magic values, such as the second magic value in the hexagonal lattice.

The flat bands inherit their topological properties from the order of the topological nodes. Existing literature primarily focuses on the physics of the two-fold degenerate Moir\'{e} flat bands. In this context, each flat band in TBG ($n=1$) possesses a Chern number of $\pm 1$ \cite{PhysRevX.10.031034}, and for the quadratic node ($n=2$) located at $M$, each flat band possesses a Chern number of $\pm 2$ \cite{Yao-Hong}. Notably, one of our findings indicates that the degeneracy can exceed two. Specifically, when discussing $n$-ordered nodes at $\Gamma$ with $n=2$ in the square lattice and $n=3$ in the hexagonal lattice, we demonstrate that by appropriately selecting half degenerate flat bands, the maximum total Chern number can be $\pm n$. Since the flat bands inherit their topology from the $n$-ordered nodes, Table II showcases these topological features in a general context. Additionally, for odd values of $n$, class DIII in 2D leads to a well-defined $\mathbb{Z}_2$ topological invariant. Due to the presence of odd Chern numbers, the $\mathbb{Z}_{2}$ topology is non-trivial.

\begin{table}[htbp]
    \begin{tabular}{*{3}{c|}*{5}{w{c}{2.5em}|}c}
        \hline\hline
        \multicolumn{3}{c|}{Order Of The Node} & 1 & 2 & 3 & 4 & 5 & $\cdots$ \\ \hline
        \multirow{4}{*}{\makecell{Square\\Lattice}} & \multirow{2}{*}{$\Gamma$} & NN & - & $4_g$ & $4_c$ & - & - & - \\ \cline{3-9}
        & & NNN & - & $2_c$ & $4_g$ & - & - & - \\ \cline{2-9}
        & \multirow{2}{*}{M} & NN & - & $2_g$ & $2_c$ & - & - & - \\ \cline{3-9}
        & & NNN & - & $2_g$ & $2_c$ & - & - & - \\ \hline
        \multirow{4}{*}{\makecell{Hexagonal\\Lattice}} & \multirow{2}{*}{$\Gamma$} & NN & - & $2_c,4_c$ & $4_g,8_g$ & $2_c,4_c$ & $4_g$ & - \\ \cline{3-9}
        & & NNN & - & $2_c,6_c$ & $4_g,8_g$ & $2_c,4_c$ & $4_g$ & - \\
        \cline{2-9}
        & \multirow{2}{*}{$\text{K}/\text{K}^{\prime}$} & NN & $2_g$ & $2_g$ & - & - & - & - \\ \cline{3-9}
        & & NNN & $2_g, 4_g$ & $2_g$ & - & - & - & - \\
        \hline
        \multicolumn{3}{c|}{\makecell{Chern Number in\\Some Flat Bands}} & $\pm 1$ & $\pm 2$ & $\pm 3$ & $\pm 4$ & $\pm 5$ & - \\
        \hline
        \multicolumn{3}{c|}{\makecell{$\mathbb{Z}_2$ Number in\\All Flat Bands}} & $1$ & $\times$ & $1$ & $\times$ & $1$ & - \\
        \hline
    \end{tabular}
    \caption{Table of the degeneracy of the MFBs for different order of the node. The numbers indicate the degeneracy of the MFBs appeared among the first four magic values, the subscript of the numbers indicates that the MFBs are gapped from or connected to other bands, denoted by 'g/c' respectively. 'NN' means only NN hopping is present. 'NNN' means besides NN hopping, NNN hopping is considered either. '-' means no MFBs appear, '$\times$' means that the $\mathbb{Z}_2$ number cannot be defined.}
    \label{MFBtable}
\end{table}

\section{The origin of the MFBs}
\label{section6}

Previously we numerically show MFBs appear at magic values. We take a step further to rigorously prove that the MFBs are exactly located at zero energy in the entire Moir\'{e} BZ. Let us go back to TBG with chiral symmetry preserved.
The key factor driving the emergence of the MFBs is that at the magic angles the zero-energy wave functions at the $\text{K}/\text{K}^{\prime}$ points exhibiting nodes in the Moir\'{e} unit cell. Using this wave function at $\text{K}$ as a base, by attaching holomorphic functions we can generate zero-energy eigenfunctions for MFBs at any momentum in the Moir\'{e} BZ \cite{Ashvin-prl19-flat}. In this study, we use the same approach to demonstrate the absolute flatness of multi-fold degenerate bands at the magic values in TBSs of the square and hexagonal lattices. To simplify this problem, we focus on solving $D_{\mathbf{k}}(\mathbf{r}) \psi_{\mathbf{k}}(\mathbf{r})=0$ for all $\mathbf{k}$. We numerically obtain the two-component wave functions $\psi_{\mathbf{M}}(\mathbf{r})$ pinned at zero energy at $\mathbf{M}$, which serves as a starting point to construct wave functions of the MFBs for other $\textbf{k}$. The conjectural wave function at $\mathbf{k}\neq\mathbf{M}$ can be expressed as $\psi_{\mathbf{k}}(\mathbf{r})\equiv f_{\mathbf{k}}(z) \psi_{\mathbf{M}}(\mathbf{r})$ with $z=x+iy$ since $D_{\mathbf{k}}(\mathbf{r})\psi_{\mathbf{k}}(\mathbf{k})=f_{\mathbf{k}}(z)[D_{\mathbf{k}}(\mathbf{r})\psi_{\mathbf{M}}(\mathbf{r})]=0$. The holomorphic function $f_{\mathbf{k}}(z)$ is either a constant or unbounded by Liouville's theorem, signifying that $f_{\mathbf{k}}(z)$ is meromorphic and thus has poles as $\mathbf{k}\neq\mathbf{M}$. The eigenfunctions $\psi_{\mathbf{k}}(\mathbf{r})$ are valid when the poles of $f_{\mathbf{k}}(z)$ are smoothed out by the nodes of $\psi_{\mathbf{M}}(\mathbf{r})$ in the Moir\'{e} unit cell. Here, we present the analytical expression for wave function of double MFBs, while further details on other multi-fold degenerate bands are provided in the supplementary materials Sec.VI.

To systematically show the band flatness, we investigate the first magic value in both square lattice and hexagonal lattice each. Specifically, we choose the quadratic node in square lattice at $\alpha^{\prime s}_{1,1}=0.197$ with the NNN hopping and the quadratic node in hexagonal lattice at $\alpha^h_{1,1}=0.156$ as shown in Fig. \ref{fig5}(f) and Fig. \ref{fig7}(b) respectively.
To distinguish the degenerate bands, we make slight shifts in the values to $\alpha^{\prime}_{s}=0.2$ and $\alpha^h_1=0.15$, presenting the corresponding band structures in Fig. \ref{fig5}(e) and Fig. \ref{fig7}(a). These two MFBs are labeled as No.1 and No. 2, and are connected to each other by the chiral symmetry operator. In our subsequent analysis, we focus on the wave function of MFB No.1, while the wave function for MFB No.2 can be obtained by applying the chiral symmetry operator to the wave function of MFB No.1. 

\begin{figure}
\centerline{\includegraphics[width=0.45\textwidth]{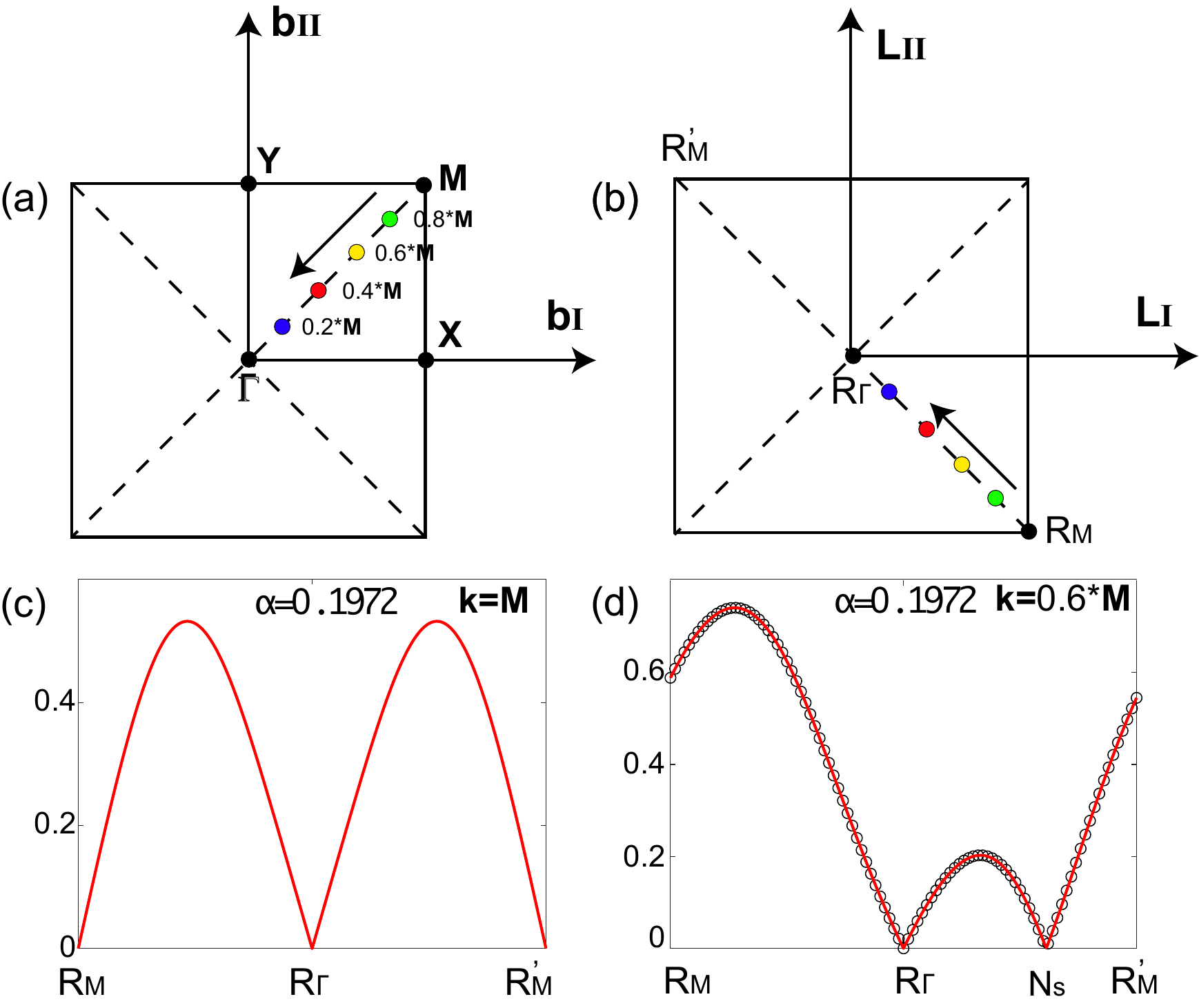}}
\caption{(a) and (b) Moir\'{e} BZ and Moir\'{e} unit cell in the square lattice. The colored dots in (a) denote the momentum we consider, the colored dots in (b) shows the nodal point of the wavefunction with the momentum having the same color in (a). (c) and (d) The red lines shows the norm of the 2-component wave functions $\psi_{\mathbf{k},1}(\mathbf{r})$ at $\mathbf{k}=\mathbf{M}$ and $\mathbf{k}=0.6\mathbf{M}$ for MFB No.1 in Moir\'{e} unit cell at the magic value $\alpha^{\prime s}_{1,1}=0.197$. '$\text{N}_{\text{s}}$' denotes the moving node. The black dots in (d) are the norm of the two-component wavefunction obtained from Eq. \ref{s_wave function1}, which exactly coincides with the numerical value.}
\label{fig9}
\end{figure}

\subsection{Origin of double MFBs in the square lattice}

Here, we show that at first magic value $\alpha^{\prime s}_{1,1}=0.197$ in the square lattice the wave function of the double MFBs are exactly zero-energy states in the Moir\'{e} BZ. Our analysis begins by using Fig. \ref{fig9}(c), which illustrates the norm of the wave function $\psi^s_{\mathbf{M},1}(\mathbf{r})$ for MFB No.$1$. There are two nodal points with a linear real-space dependence, which are situated at $\text{R}_{\Gamma}$ and $\text{R}_{\text{M}}$ in Fig. \ref{fig9}(b). Our numerical calculations reveal that the node at $\text{R}_{\text{M}}$ moves with varying momentum $\mathbf{k}$ (labeled as '$\text{N}_{\text{s}}$' in Fig. \ref{fig9}(d)), while the node at $\text{R}_{\Gamma}$ remains stationary. To create the movement of the $\psi^s_{\mathbf{k},1}(\mathbf{r})$ node, we introduce a theta function in the denominator of $f_{\textbf{k}}(z)$ eliminating the node at $\text{R}_{\text{M}}$ and another function in the numerator creating a new node.
Moreover, we adjust the parameter values of the theta functions in order to ensure the two-component wave function $\psi^s_{\mathbf{k}}(\mathbf{r})$ obeying the Bloch boundary conditions on Moir\'{e} lattice vectors $\mathbf{L}_{\text{I/II}}$, namely $\psi^s_{\mathbf{k}}(\mathbf{r}+\mathbf{L}_{\text{I/II}})=e^{i \mathbf{k}\cdot\mathbf{L}_{\text{I/II}}} \psi^s_{\mathbf{k}}(\mathbf{r})$. Finally, the analytical expression for wave function $\Psi^s_{\mathbf{k},1}(\mathbf{r})$ of MFB No.1 is given by
\begin{equation}
    \Psi^s_{\mathbf{k},1}(\mathbf{r})=\frac{\vartheta_{a, b}(\nu |\omega)}{\vartheta_{\frac{1}{2}, \frac{1}{2}}(\nu | \omega)} \Psi^s_{\mathbf{M},1}(\mathbf{r}),
    \label{s_wave function1}
\end{equation}
with $\nu={z}/{{L}_\text{I}}$, $\omega= \frac{{L}_\text{II}}{{L}_\text{I}}$ and ${L}_{\text{I/II}}=(\mathbf{L}_{\textbf{I/II}})_x+i(\mathbf{L}_{\textbf{I/II}})_y$. To satisfy the Bloch boundary conditions, the rational characteristics $a$ and $b$ must satisfy
\begin{equation}
\begin{aligned}
    a =& n_{\text{I}}+\frac{1}{2}-\frac{(\mathbf{k}-\mathbf{M}) \cdot \mathbf{L}_{\text{I}}}{2 \pi}, \\
    b =& n_{\text{II}}+\frac{1}{2}+\frac{(\mathbf{k}-\mathbf{M}) \cdot \mathbf{L}_{\text{II}}}{2 \pi},
\end{aligned}
\label{s_parameters}
\end{equation}
where $n_{\text{I}}$ and $n_{\text{II}}$ are arbitrary integers due to the lattice transnational symmetry. The definition of the theta function $\vartheta_{a, b}(\nu | \omega)$ and the detailed derivations of Eq. \ref{s_wave function1} are provided in the supplementary materials Sec.V. By definition, $f(z)={\vartheta_{a, b}}/{\vartheta_{\frac{1}{2}, \frac{1}{2}}}$ is a holomorphic function and the wave function can be extended to the entire Moir\'{e} BZ with zero energy. This is a proof for the emergence of the MFBs.

We can check if the numerical calculation is consistent with this analytic wave function. The nodal location of $\Psi_{\mathbf{k},1}(\mathbf{r})$ is denoted as  $l\mathbf{L}_{\textbf{I}}+m\mathbf{L}_{\textbf{II}}$ in the Moir\'{e} unit cell, abbreviated as ($l$,$m$). Since the node of the $\vartheta_{a, b}(\nu | \omega)$ function is at $(\frac{1}{2}-b, \frac{1}{2}-a)$, we can deduce that $l=\frac{1}{2}-b$ and $m=\frac{1}{2}-a$. With $\mathbf{k}$ varying from $\mathbf{M}$ to $\Gamma$, the nodal point moves continuously from $\text{R}_{\text{M}}$ to $\text{R}_{\Gamma}$, as demonstrated by the arrows in Fig. \ref{fig9} (a) and (b). The node locations ($l$,$m$) from the analysis coincide with the ones from the numerical solutions. In addition, the wave functions from Eq. \ref{s_wave function1} and the numerics are exactly identical as shown in Fig. \ref{fig9} (d).

From Eq. \ref{s_parameters} and Fig. \ref{fig9} we can find that, when the momentum points stay on the $\Gamma-\text{M}$ line, the zeros of the corresponding momentum stay on the $\text{R}_{\Gamma}-\text{R}_{\text{M}}$ line, which is $\frac{\pi}{2}$ rotated relative to the momentum space trajectory. This result can also be comprehended from symmetry analysis, that is, when the state in momentum space is symmetric to the $\Gamma-\text{M}$ line, the wavefunction in real space must be symmetric to the $\text{R}_{\Gamma}-\text{R}_{\text{M}}$ line, which is $\frac{\pi}{2}$ rotated relative to the momentum space mirror line. So if there is only one zero point, it must stay on the mirror line. The detailed derivation can be found in the supplementary materials Sec.IV.

\begin{figure}
\centerline{\includegraphics[width=0.45\textwidth]{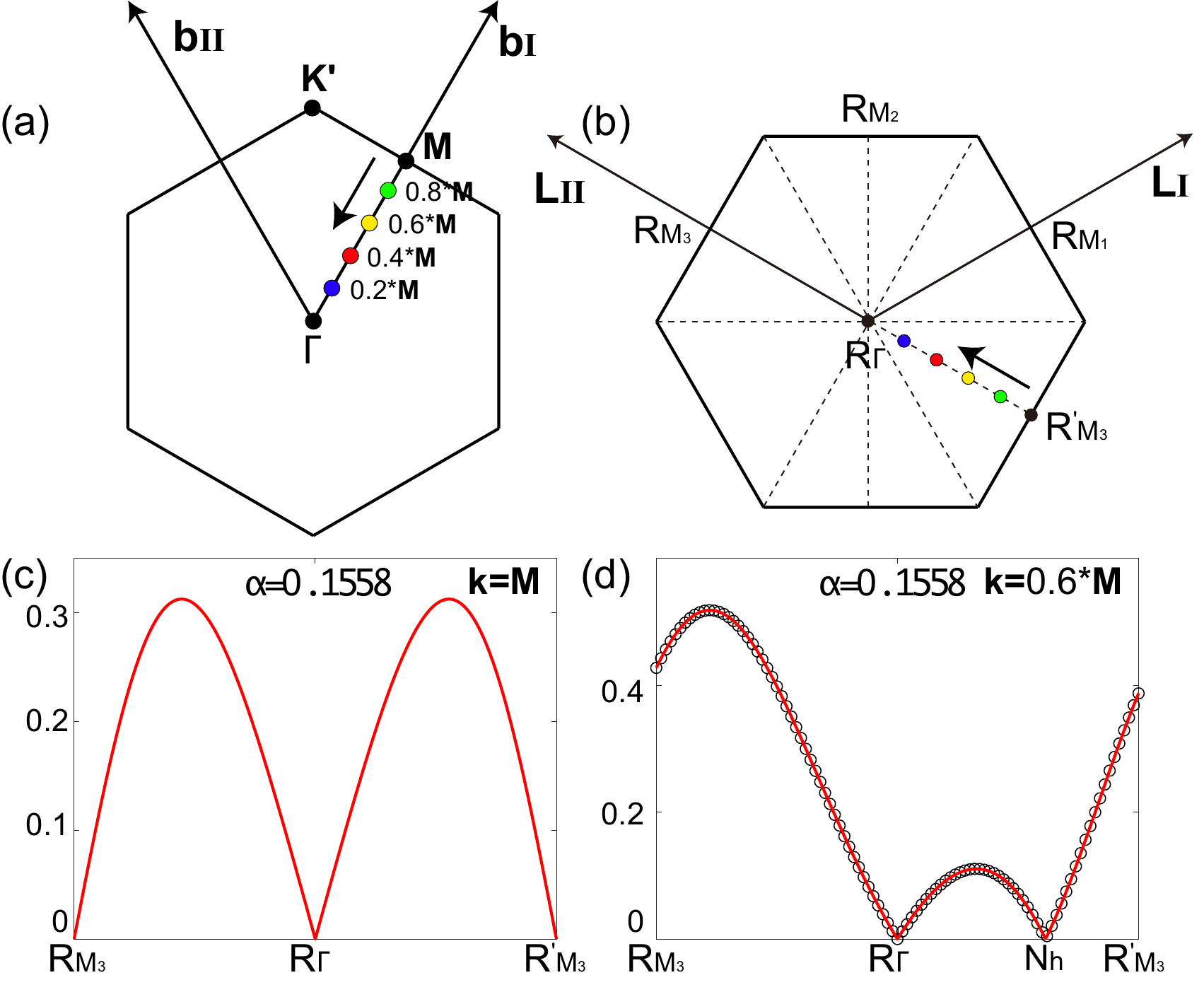}}
\caption{(a) and (b) Moir\'{e} BZ and Moir\'{e} unit cell in the hexagonal lattice. The colored dots in (a) denote the momentum we consider, the colored dots in (b) shows the nodal point of the wavefunction with the momentum having the same color in (a). (c) and (d) The red lines shows the norm of the 2-component wave functions $\psi_{\mathbf{k},1}(\mathbf{r})$ at $\mathbf{k}=\mathbf{M}$ and $\mathbf{k}=0.6\mathbf{M}$ for MFB No.1 in Moir\'{e} unit cell at the magic value $\alpha^h_{1,1}=0.156$, '$\text{N}_{\text{h}}$' denotes the moving node. The blacd circles show the norm of the wavefunction obtaind from Eq. \ref{t_wave function1}, which exactly coincides with the numerical value.}
\label{fig10}
\end{figure}

\subsection{Origin of double MFBs in the hexagonal lattice}

Here, we show that at the first magic value $\alpha^h_{1,1}=0.156$ in 
the hexagonal lattice with quadratic node, the wave function of the two-fold degenerate MFBs are exactly zero-energy states in the Moir\'e BZ.

Similarly we begin by observing the norm of the wave function $\psi^h_{\mathbf{M},1}(\mathbf{r})$ for MFB No.$1$ at the $\mathbf{M}$ point, as shown in Fig. \ref{fig10} (c), which exhibits two linear nodal points located at $\text{R}_{\Gamma}$ and $\text{R}^{\prime}_{\text{M}_3}$ in the real space Moir\'e unit cell. Our numerical calculations reveal that the node at $\text{R}^{\prime}_{\text{M}_3}$ would move as the momentum $\mathbf{k}$ varies (denoted as '$\text{N}_{\text{h}}$' in Fig. \ref{fig10}(d)), while the node at $\text{R}_{\Gamma}$ remains stationary.
So again we introduce a theta function in the numerator of $f_{\mathbf{k}}(z)$ to create the moving node of $\psi_{\mathbf{k},1}^h(\mathbf{r})$, and another theta function in the denominator to cancel the node of $\psi^h_{\mathbf{M},1}(\mathbf{r})$. We can adjust the parameters of the theta wavefunctions to ensure 
that the two-component wave function $\Psi^h_{\mathbf{k}}(\mathbf{r})$ obeys the Bloch boundary conditions on the Moir\'{e} lattice vectors $\mathbf{L}_{\text{I/II}}$, namely $\psi^h_{\mathbf{k}}(\mathbf{r}+\mathbf{L}_{\text{I/II}})=e^{i \mathbf{k}\cdot\mathbf{L}_{\text{I/II}}} \psi^t_{\mathbf{k}}(\mathbf{r})$. Finally, the analytical expression for the wave function $\Psi^h_{\mathbf{k},1}(\mathbf{r})$ of MFB No.1 is given by:
\begin{equation}
    \Psi^h_{\mathbf{k},1}(\mathbf{r})=\frac{\vartheta_{a, b}(\nu |\omega)}{\vartheta_{0, \frac{1}{2}}(\nu | \omega)} \Psi^h_{\mathbf{M},1}(\mathbf{r}) ,
    \label{t_wave function1}
\end{equation}
where $\nu={z}/{{L}_\text{I}}$, $\omega= \frac{{L}_\text{II}}{{L}_\text{I}}$ and ${L}_{\text{I/II}}=(\mathbf{L}_{\textbf{I/II}})_x+i(\mathbf{L}_{\textbf{I/II}})_y$. The rational characteristics $a$ and $b$ that satisfy the Bloch boundary conditions are given by:
\begin{equation}
    a=n_{\text{I}}+\frac{(\mathbf{k}-\mathbf{M}) \cdot \mathbf{L}_{\text{I}}}{2 \pi},
    b=n_{\text{II}}+\frac{1}{2}-\frac{(\mathbf{k}-\mathbf{M}) \cdot \mathbf{L}_{\text{II}}}{2 \pi} .
    \label{t_parameters}
\end{equation}
where $n_{\text{I}}$ and $n_{\text{II}}$ are arbitrary integers due to the lattice transnational symmetry. By definition, $f_{\mathbf{k}}(z)$ is a holomorphic function and the wave function can be extended to the entire Moir\'e BZ with zero energy.

As the momentum $\mathbf{k}$ varies in the Moir\'e BZ, the nodal point of $\psi^h_{\mathbf{M},1}(\mathbf{r})$ moves correspondingly in real space. Specifically, with $\mathbf{k}$ varying from $\text{M}$ to $\Gamma$, the nodal point moves continuously from $\text{R}^{\prime}_{\text{M}_3}$ to $\text{R}_{\Gamma}$, as demonstrated by the arrows in Fig. \ref{fig10}(a) and (b). For example, taking $\mathbf{k}=0.6\mathbf{M}$, the moving node is located at $N_t$ with ($l=0$,$m=-0.3$), which agrees with the analytical expressions Eq. \ref{t_parameters}. In addition,
the wave functions from Eq. \ref{t_wave function1} and the numerics are exactly
identical as shown in Fig. \ref{fig10} (d).

As we can find in Eq. \ref{t_parameters} and Fig. \ref{fig10}, the trajectory of the momentum points and the zeros in real space has the similar $\frac{\pi}{2}$ rotation as in the square lattice. We can apply similar symmetry analysis as in the square lattice, the detailed derivation can be found in the supplementary materials Sec.IV.

\section{Conclusion}
\label{section7}

The emergence of absolutely flat bands in TBG is rooted in the evolution of Dirac cones through the interlayer coupling tuned by twisted angles. We extend our investigation beyond linear Dirac nodes by considering higher-ordered topological nodes at various locations within the BZ. In order to search for flat bands, the locations of topological nodes can be classified into $\Gamma, \ \text{M}$ points in the square lattice and $\Gamma,\ \text{K},\ \text{K}^{\prime}$ points in the hexagonal lattice. Specifically, we find that quadratic and cubic nodes at the $\Gamma$ point in both square and hexagonal lattices can lead to MFBs at zero energy across the entire Moir\'{e} BZ. Hexagonal lattices with quartic and quintic nodes at the $\Gamma$ point also exhibit flat bands. Interestingly, for odd-ordered nodes, these flat bands exist within energy gaps, offering rich platforms for exploring strongly correlated phenomena. Furthermore, the square lattice with a quadratic or cubic node at $\text{M}$ point, and the hexagonal lattice with a linear or quadratic node at $\text{K},\ \text{K}^{\prime}$ point can also exhibit MFBs. the linear node at $\text{K},\ \text{K}^{\prime}$ point case is identical to the TBG.

TBSs include $n$-ordered topological nodes in their base layers, described by a $2\times 2$ off-diagonal Hamiltonian with a directional-dependent $n$-phase winding. After assuming that the interlayer coupling retains the same directional dependence as the topological nodes, we note that the detailed hopping form of the interlayer coupling has no impact on the criteria for flat band emergence; rather, the flat bands are primarily determined by the orders and locations of the topological nodes. In our analysis, taking into account nearest-neighbor momentum hopping suffices for predicting the emergence of MFBs. 

To identify conditions for band flatness, we compute the spectrum of the Birman-Schwinger operator's inverse. Real eigenvalues in this spectrum correspond to magic values (normalized angles) leading to flat bands. Since a state of the spectrum can be extended to two zero-energy states of the bilayer Hamiltonian, the degeneracy number of the flat bands is precisely twice the degeneracy number of the operator spectrum.
For cubic and quintic nodes in the hexagonal lattice the degeneracy number of the flat bands is invariably a multiple of four due to space-time inversion symmetry. In addition, the non-trivial topology of the MFBs inherits from the non-zero orders of the topological nodes. In many scenarios, such as quadratic and cubic nodes at $\Gamma$ in square and hexagonal lattices, some magic values yield flat bands with degeneracy greater than two-fold. The high degeneracy with the delocalization from the non-trivial topology might potentially amplify strongly correlated effects \cite{10.1038/s42254-022-00466-y}.

Strictly speaking, the Moir\'{e} spectrum plots are not enough to show band flatness in the entire Moir\'{e} BZ, since  the spectra are shown only along high symmetry lines. To fix this, we attach holomorphic functions on the state at zero energy and adjust the parameters of the holomorphic functions to satisfy the periodic boundary condition of the superlattice. This extended state is an eigenstate of the twisted bilayer Hamiltonian with zero energy and the flatness covers the entire Moir\'{e} BZ.

The development of topological nodal semimetals and topological superconductors opens up several promising avenues for band flatness in twisted bilayers. Recent discoveries include quadratic-node semimetals in photonic ring lattices \cite{photonic_lattice}, where the quadratic node benefits from symmetry protection. The bilayer of these semimetals can simply form a twisted platform with quadratic semimetals. Additionally, time-reversal symmetric topological superconductors \cite{PhysRevLett.102.187001,PhysRevLett.103.235301} naturally preserve chiral symmetry. In the case of non-trivial topological superconductors with a 3D winding number greater than one, stable high-order nodes can appear on the superconductor surface. Intriguingly, the interface between the two surfaces with the opposite 3D winding numbers can lead to twisted bilayer platforms hosting high-ordered nodes. At magic twisted angles, these two types of twisted platforms can give rise to MFBs. In particular, interesting correlated physics might emerge in these flat bands with superconductivity.

{\it Acknowledgement:}
We thank Zhida Song and Simon Becker for the helpful discussions and comments. J.H. is supported by the Ministry of Science and Technology  (Grant No. 2022YFA1403901), the National Natural Science Foundation of China (Grant No. NSFC-11888101), the Strategic Priority Research Program of the Chinese Academy of Sciences (Grant No. XDB28000000, XDB33000000), the New Cornerstone Investigator Program. X.W. is supported by the National Natural Science Foundation of China (Grant No. 12047503). C.-K.C. was supported by JST Presto Grant No.~JPMJPR2357.

\bibliography{ref}
\bibliographystyle{apsrev4-1}

\appendix
\begin{widetext}

\section{The Intetlayer Hopping Matrix}

In this section we derive the interlayer hopping matrix in detail.

The interlayer hopping matrix in $\mb{k}$-space is:
\begin{equation}
    T_{\mb{k}^t, \mb{k}^b}^{\alpha, \beta}=\bra{\mb{k}^t+\mb{H}^t, \alpha}\hat{H} \ket{\mb{k}^b+\mb{H}^b, \beta},
\end{equation}
where $\mb{k}^l+\mb{H}^l$ represents the lattice momentum of the states in the top/bottom layer,$\mb{k}^l$ is a small quantity expanded from the high symmetry point $\mb{H}$, and $\alpha/\beta$ is spin index. The Bloch states can be expanded as:
\begin{equation}
\begin{aligned}
    \ket{\mb{k}^t+\mb{H}^t, \alpha} =&\frac{1}{\sqrt{N}} \sum_{n} e^{i (\mb{k}^t+\mb{H}^t) \cdot \mb{R}^t}\ket{\mb{R}^t, \alpha}
    \\
    \ket{\mb{k}^b+\mb{H}^b, \beta} =&\frac{1}{\sqrt{N}} \sum_{n^{\prime}} e^{i (\mb{k}^b+\mb{H}^b) \cdot\mb{R}^b}\ket{\mb{R}^b, \beta},
\end{aligned}
\end{equation}
where $\mb{R}^t$ and $\mb{R}^b$ represent a real space lattice vector from the top and the bottom layer respectively. The above equations give:
\begin{equation}
    T_{\mb{k}^t, \mb{k}^b}^{\alpha, \beta} = \frac{1}{N} \sum_{\mb{R}^t, \mb{R}^b} e^{-i(\mb{k}^t+\mb{H}^t) \cdot\mb{R}^t} e^{i(\mb{k}^b+\mb{H}^b) \cdot\mb{R}^b}  \braket{\mb{R}^t, \alpha }{ \hat{H} | \mb{R}^b, \beta}.
\end{equation}
The Fourier transform of the real space interlayer hopping is:
\begin{align}
    t_{\alpha\beta}(\mb{R}^t-\mb{R}^b)=\frac{1}{4 \pi^{2}} \int d \mb{p}~\tilde{t}_{\alpha\beta}(\mb{p}) e^{i \mb{p} \cdot(\mb{R}^t-\mb{R}^b)},
\end{align}
so that
\begin{equation}
\begin{aligned}
    T_{\mb{k}^t, \mb{k}^b}^{\alpha,\beta} &= \frac{1}{N}\sum_{\mb{R}^t, \mb{R}^b} e^{-i(\mb{k}^t+\mb{H}^t) \cdot\mb{R}^t} e^{i(\mb{k}^b+\mb{H}^b) \cdot\mb{R}^b} \left[ \frac{1}{4 \pi^{2}} \int d \mb{p}~\tilde{t}_{\alpha\beta}(\mb{p}) e^{i \mb{p} \cdot(\mb{R}^t-\mb{R}^b)} \right],~~\frac{A_{\text {total }}}{(2 \pi)^{2}} \int_{\mathbb{R}^{2}} \dd \mb{p} \rightarrow  \sum_{\mb{p}}
    \\
    &= \frac{1}{N}\frac{1}{A_{\text {total }}}\sum_{\mb{R}^t, \mb{R}^b,\mb{p}} e^{-i(\mb{k}^t+\mb{H}^t) \cdot\mb{R}^t} e^{i(\mb{k}^b+\mb{H}^b) \cdot\mb{R}^b} \tilde{t}_{\alpha\beta}(\mb{p}) e^{i \mb{p} \cdot(\mb{R}^t-\mb{R}^b)}
    \\
    &= \frac{1}{A_{\text {u.c.}}}\sum_{\mb{p}} \tilde{t}_{\alpha\beta}(\mb{p}) \frac{1}{N}\sum_{\mb{R}^t} e^{-i(\mb{k}^t+\mb{H}^t-\mb{p}) \cdot\mb{R}^t} \frac{1}{N}\sum_{\mb{R}^b}e^{i(\mb{k}^b+\mb{H}^b-\mb{p}) \cdot\mb{R}^b} 
    \\
    &= \frac{1}{A_{\text {u.c.}}}\sum_{\mb{p}} \tilde{t}_{\alpha\beta}(\mb{p})\delta_{\mb{p}-(\mb{k}^t+\mb{H}^t),\mb{b}^t} \delta_{\mb{p}-(\mb{k}^b+\mb{H}^b),\mb{b}^b}
    \\
    &= \frac{\tilde{t}_{\alpha\beta}(\mb{k}^t+\mb{H}^t+\mb{b}^t)}{A_{u.c.}} \delta_{\mb{k}^t+\mb{H}^t+\mb{b}^t,\mb{k}^b+\mb{H}^b+\mb{b}^b},
    \label{intehop_general}
\end{aligned}
\end{equation}
where $\mb{b}^t$ and $\mb{b}^b$ are reciprocal vectors of the top and bottom layer, $A_{u.c.}$ is the aera of a unit cell, $A_{total}$ is the aera of the whole lattice. The delta function gives the allowed momentum difference of the interlayer hopping process
\begin{equation}
    \mb{k}^t=\mb{k}^b+(\mb{H}^b+\mb{b}^b)-(\mb{H}^t+\mb{b}^t)\equiv\mb{k}^b+\delta\mb{b}.
    \label{momentum_diff}
\end{equation}
Since the magnitude of $\mb{k}^l$ and the twist angle is small, we can approximately take $\tilde{t}(\mb{k}^t+\mb{H}^t+\mb{b}^t) \approx \tilde{t}_{\alpha\beta}(\mb{H}+\mb{b})$, then Eq. \ref{intehop_general} becomes
\begin{equation}
    T^{\alpha,\beta}_{\mb{k}^t,\mb{k}^b} = T^{\alpha,\beta}_{\mb{k}^b+\delta\mb{b},\mb{k}^b} = \frac{\tilde{t}_{\alpha\beta}(\mb{H}+\mb{b})}{A_{u.c.}} = T^{\alpha\beta}(\delta\mb{b}),
    \label{intehop_approx}
\end{equation}
where the vectors without superscript $t/b$ are untwisted. 

Now we focus on the Fourier coefficient. Suppose the interlayer hopping in real space is
\begin{equation}
\begin{aligned}
    t(\mb{r}) =& t(r)U^\dagger_{\theta_{\mb{r}}}\sigma U_{\theta_{\mb{r}}}\\
    =& t(r)e^{-i\frac{n}{2}\theta_{\mb{r}}\sigma_z} (a\sigma_x+b\sigma_y+c\sigma_z+d\sigma_0) e^{i\frac{n}{2}\theta_{\mb{r}}\sigma_z}\\
    =& t(r)
    \begin{pmatrix}
        d+c & (a-bi)e^{-in\theta_{\mb{r}}} \\
        (a+bi)e^{in\theta_{\mb{r}}} & d-c
    \end{pmatrix},
\end{aligned}
\end{equation}
where $\sigma=a\sigma_x+b\sigma_y+c\sigma_z+d\sigma_0$ is the hopping matrix in the $x$-direction. 

When $\mb{p}=\mb{0}$, the interlayer hopping matrix becomes
\begin{equation}
\tilde{t}(\mb{0})
= 
\begin{pmatrix}
    (d+c)E & (a-bi)F \\
    (a+bi)F & (d-c)E
\end{pmatrix} 
= 
\begin{pmatrix}
    (d+c)E &  \\
    & (d-c)E
\end{pmatrix},
\end{equation}
where
\begin{align}
    E &= \int t(r) r \dd r \dd\theta\\
    F &= \int t(r) e^{i n\theta} r\dd r \dd\theta=\int  t(r) e^{-i n\theta} r\dd r \dd\theta=0 .
\end{align}

When $\mb{p}\neq \mb{0}$, the hopping matrix elements 
\begin{equation}
\begin{aligned}
    \tilde{t}_{12}(\mb{p}) 
    &= \int e^{-i \mb{p} \cdot \mb{r}} t(r) (a-bi)e^{-i n\theta_{\mb{r}}} \dd \mb{r} 
    \\
    &= (a-bi)\int e^{-ipr\cos(\theta_{\mb{r}}-\theta_{\mb{p}})} t(r) e^{-i n\theta_{\mb{r}}} r\dd r\dd\theta_{\mb{r}} 
    \\
    &= (a-bi)e^{-in\theta_{\mb{p}}}\int e^{-ipr\cos(\theta_{\mb{r}}-\theta_{\mb{p}})} t(r) e^{-i n(\theta_{\mb{r}}-\theta_{\mb{p}})} r\dd r\dd(\theta_{\mb{r}}-\theta_{\mb{p}})
    \\
    &=(a-bi)e^{-in\theta_{\mb{p}}}\int e^{-ipr\cos\theta} t(r) e^{i n\theta} r\dd r\dd\theta ,
\end{aligned}
\end{equation}

\begin{equation}
\begin{aligned}
    \tilde{t}_{21}(\mb{p}) 
    &= \int e^{-i \mb{p} \cdot \mb{r}} t(r) (a+bi)e^{i n\theta_{\mb{r}}} \dd \mb{r} 
    \\
    &= (a+bi)\int e^{-ipr\cos(\theta_{\mb{r}}-\theta_{\mb{p}})} t(r) e^{i n\theta_{\mb{r}}} r\dd r\dd\theta_{\mb{r}} 
    \\
    &= (a+bi)e^{in\theta_{\mb{p}}}\int e^{-ipr\cos(\theta_{\mb{r}}-\theta_{\mb{p}})} t(r) e^{i n(\theta_{\mb{r}}-\theta_{\mb{p}})} r\dd r\dd(\theta_{\mb{r}}-\theta_{\mb{p}})
    \\
    &= (a+bi)e^{in\theta_{\mb{p}}}\int e^{-ipr\cos\theta
    } t(r) e^{i n\theta} r\dd r \dd\theta ,
\end{aligned}
\end{equation}

the diagonal terms are
\begin{equation}
\begin{aligned}
    \tilde{t}_{11}(\mb{p}) 
    &= (d+c)\int e^{-i \mb{p} \cdot \mb{r}} t(r) \dd \mb{r} 
    \\
    &= (d+c)\int e^{-ipr\cos(\theta_{\mb{r}}-\theta_{\mb{p}})} t(r) r\dd r \dd(\theta_{\mb{r}}-\theta_{\mb{p}})
    \\
    &= (d+c)\int e^{-ipr\cos\theta} t(r) r\dd r \dd\theta ,
\end{aligned}
\end{equation}

\begin{equation}
\begin{aligned}
    \tilde{t}_{22}(\mb{p})
    &= (d-c)\int e^{-i \mb{p} \cdot \mb{r}} t(r) \dd \mb{r} 
    \\
    &= (d-c)\int e^{-ipr\cos(\theta_{\mb{r}}-\theta_{\mb{p}})} t(r) r\dd r \dd(\theta_{\mb{r}}-\theta_{\mb{p}})
    \\
    &= (d-c)\int e^{-ipr\cos\theta} t(r) r \dd r \dd\theta ,
\end{aligned}
\end{equation}
the hopping matrix is now
\begin{equation}
\tilde{t}(\mb{p}) = 
\begin{pmatrix}
    (d+c)D(\mb{p}) & (a-bi)C_{J_z}(\mb{p})e^{-in\theta_{\mb{p}}} \\
    (a+bi)C_{J_z}(\mb{p})e^{in\theta_{\mb{p}}} & (d-c)D(\mb{p})
\end{pmatrix} ,
\end{equation}
where
\begin{equation}
\begin{aligned}
    D(\mb{p}) &= \int e^{-ipr\cos\theta} t(r) r\dd r\dd\theta 
    \\
    C_{J_z}(\mb{p}) &= \int e^{-ipr\cos\theta} t(r) e^{i n\theta} r \dd r \dd\theta=\int e^{-ipr\cos\theta} t(r) e^{-i n\theta} r \dd r \dd\theta ,
\end{aligned}
\end{equation}
with $J_z\equiv\frac{n}{2}$. When $b=c=d=0$, the hopping matrix becomes:
\begin{equation}
\begin{aligned}
    & \tilde{t}(\mb{0})= \mb{0}_{2\times 2}
    \\
    & \tilde{t}(\mb{p}) = aC_{J_z}(\mb{p})
    \begin{pmatrix}
        & e^{-2iJ_z\theta_{\mb{p}}} \\
        e^{2iJ_z\theta_{\mb{p}}} & 
    \end{pmatrix} .
\end{aligned}
\label{intehop_Fc}
\end{equation}

\section{Hamiltonian of twisted bialyer systems}

In this section we derive the Hamiltonian of the twisted bilayer systems with a n-order node at the M point in the square lattice or at the K/K' point in the hexagonal lattice. The derivition is basically the same as in the main text, just with the interlayer hopping terms different. To be noticed, unlike in the $\Gamma$ point case, the two states having the same mometum but from different layer are not coupled together in any case, so these Hamiltonians cannot be divded into two decoupled sub-Hamitonians, and the bands are not naturally two-fold degenerate.

\subsection{square lattice with a M point node}

When the node is located at the M point, we have $\mb{H}^l=\mb{M}^l$ in Eq. \ref{momentum_diff}, that is
\begin{equation}
\begin{aligned}
    \mb{k}^t =& \mb{k}^b+\delta\mb{b} \\
    \delta\mb{b} =& (\mb{M}^t+\mb{b}^t)-(\mb{M}^b+\mb{b}^b)
\end{aligned}
\end{equation}
When $\mb{b}^l$ equals to $\mb{0},-\mb{b}_1^l,-\mb{b}_2^l,-\mb{b}_1^l-\mb{b}_2^l$, denoted as $\mb{G}_i^l$, the corresponding $\delta\mb{b}$ are the minimal ones, which are denoted as $\mb{q}_i$ in Fig. \ref{SM_fig1}(a), when $\mb{b}^l$ equals to $\mb{b}_2^l,-\mb{b}_1^l+\mb{b}_2^l,-2\mb{b}_1^l,-2\mb{b}_1^l-\mb{b}_2^l,-\mb{b}_1^l-2\mb{b}_2^l,-2\mb{b}_2^l,\mb{b}_1^l-\mb{b}_2^l,\mb{b}_1^l$, denoted as $\tilde{\mb{G}}_j^l$, the corresponding $\delta\mb{b}$ are the second minimal ones, which are denoted as $\tilde{\mb{q}}_j$ in Fig.  \ref{SM_fig1}(a). That is,
\begin{equation}
\begin{aligned}
    \mb{q}_i=&(\mb{M}^t+\mb{G}_i^t)-(\mb{M}^b+\mb{G}^b_i)\equiv \mb{M}_i^t-\mb{M}_i^b
    \\
    \tilde{\mb{q}}_j=&(\mb{M}^t+\tilde{\mb{G}}_j^t)-(\mb{M}^b+\tilde{\mb{G}}^b_j) \equiv \tilde{\mb{M}}_j^t-\tilde{\mb{M}}_j^b
\end{aligned}
\end{equation}
\begin{figure}[H]
    \centering
    \includegraphics[scale=0.35]{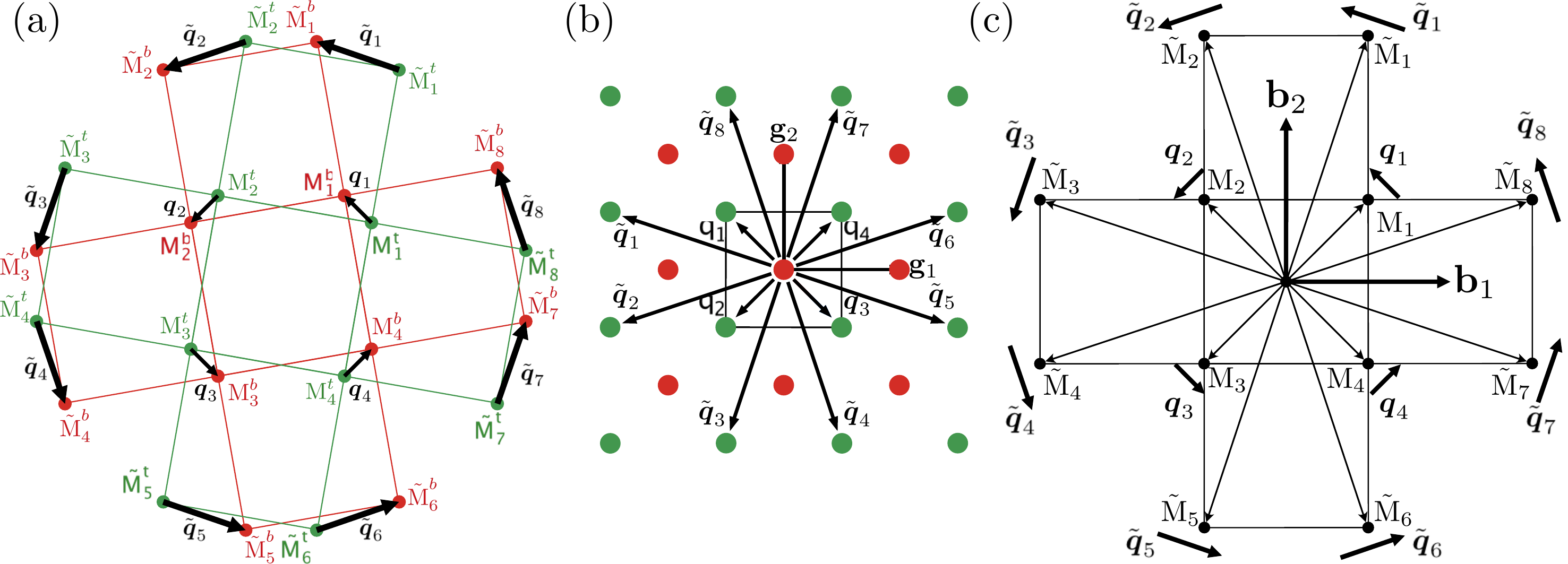}
    \caption{(a) The monolayer square BZs with the twist form moir\'{e} BZs.
    The red and green hexagonals represent the BZs of the bottom and top layers, each rotated by an angle $\pm\theta/2$ (b) The corresspounding square and hexagonalal reciprocal lattice structures. The red
    and green atoms indicate Dirac cones of the top and bottom layers, the small black square and hexagonal indicates the moir\'{e} BZ. (c) The NN and NNN high symmetry points of the monolayer BZ, the nearyby vectors are the corresponding hopping vector resulted from the twisting of the high symmetry point, see Eq. \ref{sqr_M_intehop}}
    \label{SM_fig1}
\end{figure}

The Fourier coefficient from Eq. \ref{intehop_approx} is decided by Eq. \ref{intehop_Fc}
\begin{equation}
    \tilde{t}(\mb{M}+\mb{G}_i/\tilde{\mb{G}}_j)=\tilde{t}(\mb{M}_i/\tilde{\mb{M}}_j) = aC_{J_z}(\mb{M}_i/\tilde{\mb{M}}_j)
    \begin{pmatrix}
        & e^{-2iJ_z\theta_{\mb{M}_i/\tilde{\mb{M}}_j}} \\
        e^{2iJ_z\theta_{\mb{M}_i/\tilde{\mb{M}}_j}} & 
    \end{pmatrix}
    \label{sqr_M_Tq}
\end{equation}
The interlayer hopping matrix is then
\begin{equation}
    T(\mb{q}_i/\tilde{\mb{q}}_j)= \frac{\tilde{t}(\mb{M}_i/\tilde{\mb{M}}_j)}{A_{u.c.}}
    \label{sqr_M_intehop}
\end{equation}
Comparing the interlayer hopping matrix with that of the models with a node at $\Gamma$ point in the main text, we can easily see that the two set of NN interlayer hopping matrix just differ by a $\frac{\pi}{4}$ rotation. So when only NN hopping is considered, the Hamiltonian of the two TBSs with a node at $\Gamma$ point and M point are essentially the same.

The repeated hopping forms a momentum space lattice, as shown in Fig. \ref{SM_fig1}(b). The primitive lattice vectors of this momentum lattice are just that of the Moir\'e BZ, which are denote as $\mb{g}_i$, shown in Fig. \ref{SM_fig1}(b). The size of the Moir\'e BZ can be characterized by the magnitude of the vectors $\mb{g}_i$ , which is
\begin{equation}
    |\mb{g}_i|=2|\mb{b}_i|\sin\frac{\theta}{2}
\end{equation}
Where $\mb{b}_i$ is the primitive lattice vectors of the monolayer BZ. 

Suppose the center of the momentum space lattice is $\mb{k}$ from the bottom layer, the momentum of the other lattice sites from bottom layer (denoted by the red dots in Fig. \ref{SM_fig1}(b)) can be expressed as $\mb{k}+\mb{g}$, and the lattice sites from the top layer that can hop with these sites can be expressed as $\mb{k}+\mb{g}+\mb{q}_i/\tilde{\mb{q}}_j$. Here $\mb{k}$ is defined in the Moir\'e Brillouin zone, and $\mb{g} = m\mb{g}_1 + n\mb{g}_2$ is a Moir\'e reciprocal lattice vector. Then total Hamiltonian is
\begin{equation}
    \hat{H} = \sum^{l,\alpha,\beta}_{\mb{k},\mb{g}} \hat{\psi}^{l\dagger}_{\mb{k}+\mb{g},\alpha} h^{\alpha\beta}(\mb{k}+\mb{g}) \hat{\psi}^l_{\mb{k}+\mb{g},\beta} + \sum^{\alpha,\beta}_{\mb{k},\mb{g},i,j} \hat{\psi}^{t\dagger}_{\mb{k}+\mb{g}+\mb{q}_i/\tilde{\mb{q}}_j,\alpha} T^{\alpha\beta}(\mb{q}_i/\tilde{\mb{q}}_j) \hat{\psi}^b_{\mb{k}+\mb{g},\beta}+ h.c
    \label{sqr_M_total_Hamiltonian_k}
\end{equation}

Changing to the basis $\hat{\psi}_{\mb{k},\alpha}(\mb{r})$ and reshuffling the basis, we can get the real space Hamiltonian
\begin{equation}
    \begin{aligned}
        H_{\mathbf{k},J_z}(\mathbf{r}) =
        &\begin{pmatrix}
            0 & D_{\mathbf{k},J_z}^{*}(-\mathbf{r}) \\
            D_{\mathbf{k},J_z}(\mathbf{r}) & 0
        \end{pmatrix}
        \\
        D_{\mathbf{k},J_z}(\mathbf{r}) =
        &\begin{pmatrix}
            (\bar{k}-i2\bar{\partial})^n &  \alpha_{J_z} U_{J_z}(\mathbf{r}) \\
            \alpha_{J_z} U_{J_z}(-\mathbf{r}) & (\bar{k}-i2\bar{\partial})^n
        \end{pmatrix}
    \end{aligned}
    \label{sqr_M_total_Hamiltonian_r}
\end{equation}
with
\begin{equation}
\begin{aligned}
    & U_{J_z}(\mb{r}) = U^{\text{NN}}_{J_z}(\mb{r}) + U^{\text{NNN}}_{J_z}(\mb{r})
    \\
    & U^{\text{NN}}_{J_z}(\mb{r})= \sum_{i} e^{-i\mb{q}_i\cdot\mb{r}} T_{21}(\mb{q}_i)
    \\
    & U^{\text{NNN}}_{J_z}(\mb{r})= \sum_{j} e^{-i\tilde{\mb{q}}_j\cdot\mb{r}} T_{21}(\tilde{\mb{q}}_j)
\end{aligned}
\label{real_space_intehop}
\end{equation}
Using these Hamiltonians we can proceed to calculate the spectrum of Birman-Schwinger operator and the energy bands.

\subsection{hexagonal lattice with a $\text{K}/\text{K}^\prime$ point node}

When the node is located at the $\text{K}$ point, we have $\mb{H}^l=\mb{K}^l$ in Eq. \ref{momentum_diff}, that is
\begin{equation}
\begin{aligned}
    \mb{k}^t =& \mb{k}^b+\delta\mb{b} \\
    \delta\mb{b} =& (\mb{K}^t+\mb{b}^t)-(\mb{K}^b+\mb{b}^b)
\end{aligned}
\end{equation}
When $\mb{b}^l$ equals to $\mb{0},-\mb{b}_1^l,-\mb{b}_1^l-\mb{b}_2^l,$, denoted as $\mb{G}_i^l$, the corresponding $\delta\mb{b}$ are the minimal ones, denoted as $\mb{q}_i$ in Fig. \ref{SM_fig2}(a), when $\mb{b}^l$ equals to $-\mb{b}_2^l,\mb{b}_2^l,-2\mb{b}_1^l-\mb{b}_2^l$, denoted as $\tilde{\mb{G}}_j^l$, the corresponding $\delta\mb{b}$ are the second minimal ones, which are denoted as $\tilde{\mb{q}}_j$ in Fig. \ref{SM_fig2}. That is,
\begin{equation}
\begin{aligned}
    \mb{q}_i=&(\mb{K}^t+\mb{G}_i^t)-(\mb{K}^b+\mb{G}^b_i)\equiv \mb{K}_i^t-\mb{K}_i^b
    \\
    \tilde{\mb{q}}_j=&(\mb{K}^t+\tilde{\mb{G}}_j^t)-(\mb{K}^b+\tilde{\mb{G}}^b_j) \equiv \tilde{\mb{K}}_j^t-\tilde{\mb{K}}_j^b
\end{aligned}
\end{equation}

\begin{figure}[H]
    \centering
    \includegraphics[scale=0.35]{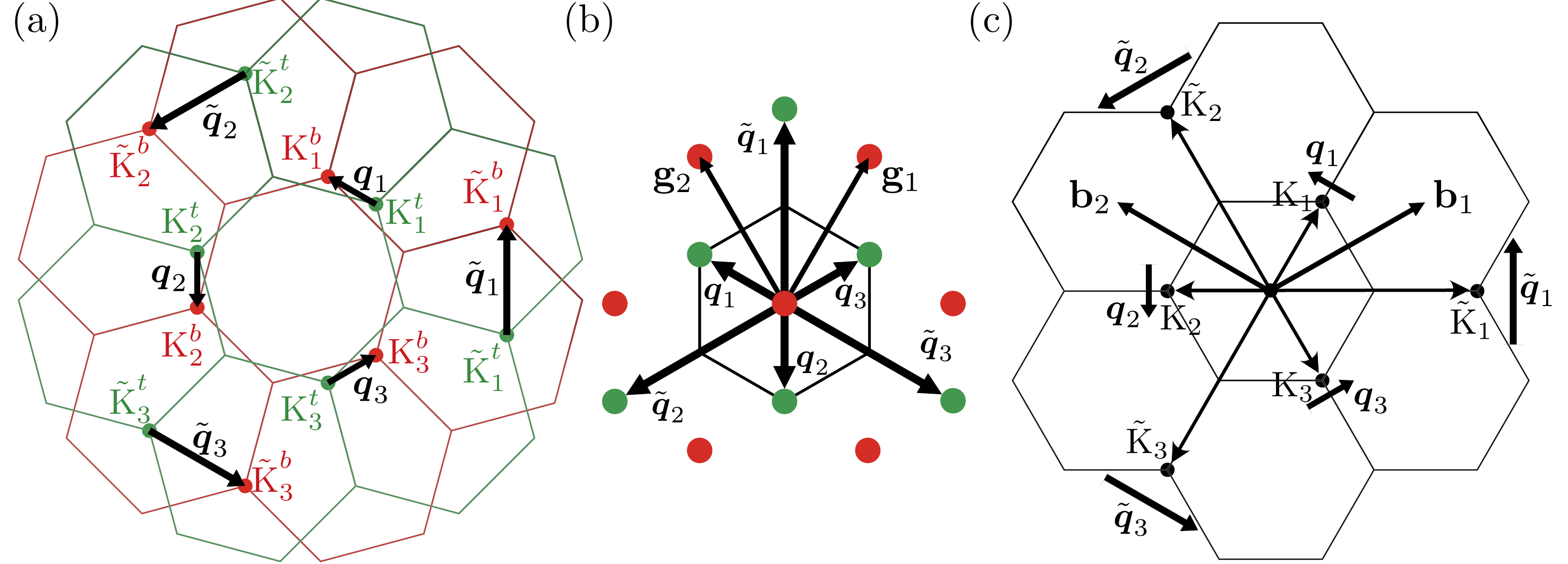}
    \caption{(a) The monolayer hexagonal BZs with the twist form moir\'{e} BZs.
    The red and green hexagonals represent the BZs of the bottom and top layers, each rotated by an angle $\pm\theta/2$ (b) The corresspounding hexagonalal reciprocal lattice structures. The red
    and green atoms indicate Dirac cones of the top and bottom layers, the small black square and hexagonal indicates the moir\'{e} BZ. (c) The NN and NNN high symmetry points of the monolayer BZ, the nearyby vectors are the corresponding hopping vector resulted from the twisting of the high symmetry point, see Eq. \ref{hex_K_intehop}}
    \label{SM_fig2}
\end{figure}

The Fourier coefficient from Eq. \ref{intehop_approx} is decided by Eq. \ref{intehop_Fc}
\begin{equation}
    \tilde{t}(\mb{K}+\mb{G}_i/\tilde{\mb{G}}_j)=\tilde{t}(\mb{K}_i/\tilde{\mb{K}}_j) = aC_{J_z}(\mb{K}_i/\tilde{\mb{K}}_j)
    \begin{pmatrix}
        & e^{-2iJ_z\theta_{\mb{K}_i/\tilde{\mb{K}}_j}} \\
        e^{2iJ_z\theta_{\mb{K}_i/\tilde{\mb{K}}_j}} & 
    \end{pmatrix}
\end{equation}
The interlayer hopping matrix is then
\begin{equation}
    T^{\alpha\beta}(\mb{q}_i/\tilde{\mb{q}}_j)= \frac{\tilde{t}(\mb{K}_i/\tilde{\mb{K}}_j)}{A_{u.c.}}
    \label{hex_K_intehop}
\end{equation}

The repeated hopping forms a momentum space lattice, as shown in Fig. \ref{SM_fig2}(b). The primitive lattice vectors of this momentum lattice are just that of the Moir\'e BZ, which are denote as $\mb{g}_i$, shown in Fig. \ref{SM_fig2}(b). Suppose the center of the momentum space lattice is $\mb{k}$ from the bottom layer, the momentum of the other lattice sites from bottom layer (denoted by the red dots in Fig. \ref{SM_fig2}(b)) can be expressed as $\mb{k}+\mb{g}$, and the lattice sites from the top layer that can hop with these sites can be expressed as $\mb{k}+\mb{g}+\mb{q}_i/\tilde{\mb{q}}_j$. Here $\mb{k}$ is defined in the Moir\'e Brillouin zone, and $\mb{g} = m\mb{g}_1 + n\mb{g}_2$ is a Moir\'e reciprocal lattice vector. The total Hamiltonian then has the same form as Eq. \ref{sqr_M_total_Hamiltonian_k}, Eq. \ref{sqr_M_total_Hamiltonian_r} and Eq. \ref{real_space_intehop}. Using these Hamiltonians we can proceed to calculate the spectrum of Birman-Schwinger operator and the energy bands.

By observing Eq. \ref{hex_K_intehop} and Fig. \ref{SM_fig2} we can see that, the $\text{K}^{\prime}$ point case and the K point case just differ by a $\frac{\pi}{3}$ rotation, so the two models are essentially the same. To be noticed, when we consider the linear case, that is, when there is a linear Dirac cone located at the $\text{K}/\text{K}^{\prime}$ point, we can see the interlayer hopping terms from Eq. \ref{hex_K_intehop} are just the same as that of the twisted bilayer graphene, and so does the Hamiltonian.

\section{Chern Number of the MFBs}

Here, we investigate the topological properties of highly degenerate flat bands in both square and hexagonal lattices. In the square lattice, Fig.\ref{SM_fig3}(b) illustrates the Wilson loop $W(\mathbf{k}_x)$ for the quadruple flat bands, as highlighted by the black dashed box in Fig.\ref{SM_fig3}(a). Nevertheless, the two occupied bands No. (1-2) possess a total Chern number of +2, as indicated by the blue points in Fig.\ref{SM_fig3} (b). In contrast, the two unoccupied bands exhibit opposing total Chern numbers, represented by the red circles in Fig.\ref{SM_fig3} (b). In the hexagonal lattice, for the cubic nodes, Fig.\ref{SM_fig3}(d) and (f) depict the Wilson loops $W(\mathbf{k}_x)$ corresponding to the four-fold and eight-fold degenerate flat bands, as highlighted by the black dotted boxes in Fig.\ref{SM_fig3} (c) and (e). This reveals that they possess a nontrivial topology $\mathbb{Z}_{2}=1$. Within both the four-fold and eight-fold degenerate flat bands, the occupied bands No. (1-2) and No. (1-4) share identical total Chern numbers of +3, as denoted by the blue and green points in Fig.\ref{SM_fig3} (d) and (f). In contrast, the unoccupied bands No. (3-4) and No. (5-8) exhibit opposite total Chern numbers, represented by the red and green circles in Fig.\ref{SM_fig3} (d) and (f).

\begin{figure}[H]
\centerline{\includegraphics[width=0.9\textwidth]{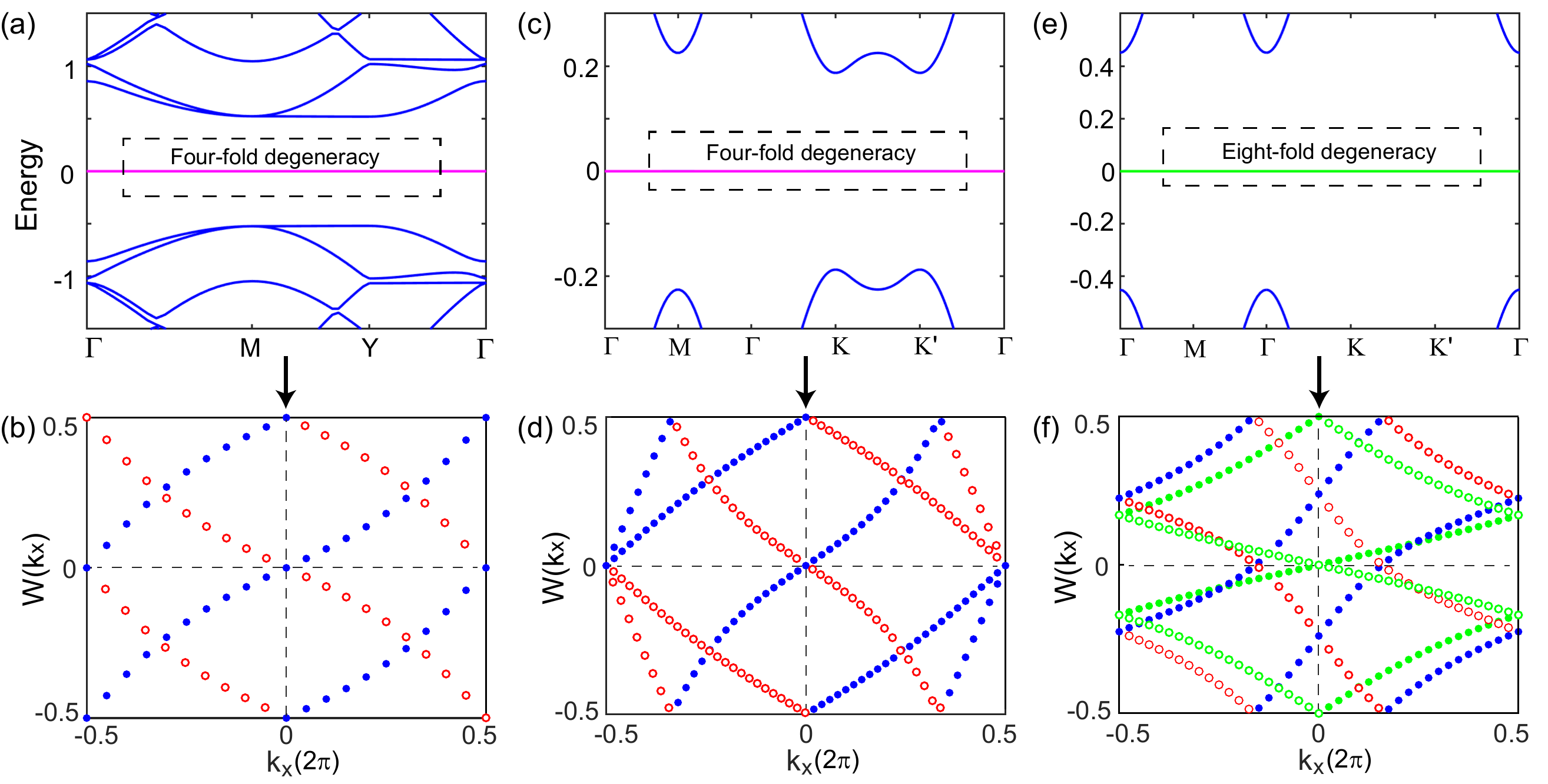}}
\caption{(color online)  Band structures and Wilson loops in the square and  hexagonal lattices. (a) and (b) Band structures and Wilson loops $W(\mathbf{k}_x)$ of quadruple flat bands at magic value $\alpha^{s}_{1,1}=0.264$ without NNN hopping term. (c)-(f) Band structures and corresponding Wilson loops of four-fold and eight-fold degenerate flat bands without NNN hopping term in the hexagonal lattice at magic values $\alpha^{h}_{\frac{3}{2},1}=0.127$ and $\alpha^{h}_{\frac{3}{2},2}=0.846$. \label{SM_fig3} }
\end{figure}

\section{Symmetries of the twisted bilayer systems}

In this section, we list some important symmetry operators of the twisted bilayer systems (TBSs). The bilayers introduce a layer degree, the monolayer rotation symmetry around z-axis is inherited by the TBSs, the rotation symmetry with an in-plane axis are different, since they exchange the layer degree.

Under the basis $\hat{\Phi}_{\mb{k}}(\mb{r})=(\hat{\phi}_{\mb{k}}^t(\mb{r}),\hat{\chi}_{\mb{k}}^t(\mb{r}),\hat{\phi}_{\mb{k}}^b(\mb{r}),\hat{\chi}_{\mb{k}}^b(\mb{r}))^{\text{T}}$, the monolayer symmetry operators are
\begin{table}[H]
    \centering
    \begin{tabular}{c|c|c}
    \hline\hline
    \text{order n} & \text{even} & \text{odd} \\
    \hline
    \text{rotation $\mathcal{R}_{\theta}$} & \multicolumn{2}{c}{$e^{i\frac{n}{2}\theta\sigma_z}$} \\
    \hline
    \text{time reversal } $\mathcal{T}$ & $\sigma_1\mathcal{K}$ & $\sigma_2\mathcal{K}$ \\
    \hline
    \text{inversion } $\mathcal{I}$ & $\sigma_0$ & $\sigma_3$ \\
    \hline
    \end{tabular}
    \label{monolayer symmetry}
    \caption{The symmetry operators of the monolayer Hamiltonian}
\end{table}
The bilayer symmetry operators are
\begin{table}[H]
    \centering
    \begin{tabular}{c|c|c}
    \hline\hline
    \text{order } n & \text{even} & \text{odd} \\
    \hline
    \text{rotation $\mathcal{R}_{\theta}$} & \multicolumn{2}{c}{$\tau_0 e^{i\frac{n}{2}\theta\sigma_z}$} \\
    \hline
    \text{time reversal } $\mathcal{T}$ & $\tau_0\sigma_1\mathcal{K}$ & $\tau_0\sigma_2\mathcal{K}$ \\
    \hline
    \text{space inversion } $\mathcal{I}$ & $\tau_1\sigma_0$ & $\tau_2\sigma_3$ \\
    \hline
    $\mathcal{I}\mathcal{T}$ & $\tau_1\sigma_1\mathcal{K}$ & $-i\tau_2\sigma_1\mathcal{K}$ \\
    \hline
    \end{tabular}
    \label{bilayer symmetry}
    \caption{The symmetry operators of the bilayer Hamiltonian}
\end{table}
Here,$\sigma_i$ and $\tau_i$ are Pauli matrix denotes the internal (spin or orbital) and layer degree of freedom. From this table we can find that, for nodes with odd order, besides the time reversal symmetry $\mathcal{T}^2=-1$, there is also the symmetry $(\mathcal{I}\mathcal{T})^2=-1$ (for models with the node located at the $\Gamma$ point, if the node is located at other points, there is no inversion symmetry), these two symmetries ensure that each energy bands are at least two-fold degenerate. 

We can also analyze the symmetric properties of the wavefunctions. Suppose the symmetry operation is denoted as $g$
\begin{equation}
\begin{aligned}
    & H_{\mb{k}}(\mb{r})\Psi_{\mb{k}}(\mb{r})=E_{\mb{k}}\Psi_{\mb{k}}(\mb{r}) &
    \Longrightarrow &
    U_g^\dagger H_{\mb{k}}(\mb{r})U_gU_g^\dagger \Psi_{\mb{k}}(\mb{r}) =E_{\mb{k}} U_g^\dagger\Psi_{\mb{k}}(\mb{r}) \\
    &&\Longrightarrow &
    H_{g\mb{k}}(g\mb{r})U_g^\dagger \Psi_{\mb{k}}(\mb{r})=E_{\mb{k}}U_g^\dagger \Psi_{\mb{k}}(\mb{r}) \\
    & H_{g\mb{k}}(g\mb{r})\Psi_{g\mb{k}}(g \mb{r})=E_{g\mb{k}} \Psi_{g\mb{k}}(g\mb{r}), & &
\end{aligned}
\end{equation}
we have
\begin{equation}
    U_g^\dagger \Psi_{\mb{k}}(\mb{r}) = \Psi_{g\mb{k}}(g \mb{r}).
\end{equation}

When we consider models with an even order node, applying $\mathcal{IT}$ to the basis
\begin{equation}
    U^{\dagger}_{\mathcal{IT}}\Psi_{\mb{k}}(\mb{r})=
    \begin{pmatrix}
        & & & 1 \\
        & & 1 &  \\
        & 1 & & \\
    1 & & & 
    \end{pmatrix}
    \mathcal{K}
    \begin{pmatrix}
    \psi^t_{\mb{k}}(\mb{r}) \\
    \chi^t_{\mb{k}}(\mb{r}) \\
    \psi^b_{\mb{k}}(\mb{r}) \\
    \chi^b_{\mb{k}}(\mb{r})
    \end{pmatrix}
    =
    \begin{pmatrix}
    \chi^{b*}_{\mb{k}}(\mb{r}) \\
    \psi^{b*}_{\mb{k}}(\mb{r}) \\
    \chi^{t*}_{\mb{k}}(\mb{r}) \\
    \psi^{t*}_{\mb{k}}(\mb{r})
    \end{pmatrix}
    =\Psi_{\mathcal{IT}\mb{k}}(\mathcal{IT}\mb{r})=
    \begin{pmatrix}
    \psi^t_{\mb{k}}(-\mb{r}) \\
    \chi^t_{\mb{k}}(-\mb{r}) \\
    \psi^b_{\mb{k}}(-\mb{r}) \\
    \chi^b_{\mb{k}}(-\mb{r})
    \end{pmatrix},
\end{equation}
so we have
\begin{equation}
    \begin{pmatrix}
    \psi^t_{\mb{k}}(\mb{r}) \\
    \chi^t_{\mb{k}}(\mb{r}) \\
    \psi^b_{\mb{k}}(\mb{r}) \\
    \chi^b_{\mb{k}}(\mb{r})
    \end{pmatrix}
    =
    \begin{pmatrix}
    \chi^{b*}_{\mb{k}}(-\mb{r}) \\
    \psi^{b*}_{\mb{k}}(-\mb{r}) \\
    \chi^{t*}_{\mb{k}}(-\mb{r}) \\
    \psi^{t*}_{\mb{k}}(-\mb{r})
    \end{pmatrix}.
    \label{IT_on_basis}
\end{equation}

In square lattice, we consider the rotation around the diagonal line $y=x$ in real space, which we denote as $\mathcal{C}_{xy}$. The representation of the rotation symmetry operator $\mathcal{C}_{xy}$ can be derived:
\begin{equation}
    U_{\mathcal{C}_{xy}}=\tau_1\sigma_2 .
\end{equation}
Applying this operator to the basis
\begin{equation}
    U^{\dagger}_{\mathcal{C}_{xy}}\Psi_{\mb{k}}(\mb{r})=
    \begin{pmatrix}
        & & & -i \\
        & & i &  \\
        & -i & & \\
    i & & & 
    \end{pmatrix}
    \begin{pmatrix}
    \psi^t_{\mb{k}}(\mb{r}) \\
    \chi^t_{\mb{k}}(\mb{r}) \\
    \psi^b_{\mb{k}}(\mb{r}) \\
    \chi^b_{\mb{k}}(\mb{r})
    \end{pmatrix}
    =
    \begin{pmatrix}
    -i\chi^b_{\mb{k}}(\mb{r}) \\
    i\psi^b_{\mb{k}}(\mb{r}) \\
    -i\chi^t_{\mb{k}}(\mb{r}) \\
    i\psi^t_{\mb{k}}(\mb{r})
    \end{pmatrix}
    =\Psi_{\mathcal{C}_{xy}\mb{k}}(\mathcal{C}_{xy}\mb{r})=
    \begin{pmatrix}
    \psi^t_{\mathcal{C}_{xy}\mb{k}}(\mathcal{C}_{xy}\mb{r}) \\
    \chi^t_{\mathcal{C}_{xy}\mb{k}}(\mathcal{C}_{xy}\mb{r}) \\
    \psi^b_{\mathcal{C}_{xy}\mb{k}}(\mathcal{C}_{xy}\mb{r}) \\
    \chi^b_{\mathcal{C}_{xy}\mb{k}}(\mathcal{C}_{xy}\mb{r})
    \end{pmatrix},
\end{equation}
so that
\begin{equation}
    \begin{pmatrix}
        \chi^b_{\mb{k}}(\mb{r}) \\
        \psi^b_{\mb{k}}(\mb{r}) \\
        \chi^t_{\mb{k}}(\mb{r}) \\
        \psi^t_{\mb{k}}(\mb{r})
    \end{pmatrix}
    =
    \begin{pmatrix}
    i\psi^t_{\mathcal{C}_{xy}\mb{k}}(\mathcal{C}_{xy}\mb{r}) \\
    -i\chi^t_{\mathcal{C}_{xy}\mb{k}}(\mathcal{C}_{xy}\mb{r}) \\
    i\psi^b_{\mathcal{C}_{xy}\mb{k}}(\mathcal{C}_{xy}\mb{r}) \\
    -i\chi^b_{\mathcal{C}_{xy}\mb{k}}(\mathcal{C}_{xy}\mb{r})
    \end{pmatrix}.
    \label{Cxy_on_basis}
\end{equation}
Using Eq. \ref{Cxy_on_basis} together with Eq. \ref{IT_on_basis}, we have
\begin{equation}
    \begin{pmatrix}
    \psi^t_{\mb{k}}(\mb{r}) \\
    \chi^t_{\mb{k}}(\mb{r}) \\
    \psi^b_{\mb{k}}(\mb{r}) \\
    \chi^b_{\mb{k}}(\mb{r})
    \end{pmatrix}
    =
    \begin{pmatrix}
    i\psi^{t*}_{\mathcal{C}_{xy}\mb{k}}(-\mathcal{C}_{xy}\mb{r}) \\
    -i\chi^{t*}_{\mathcal{C}_{xy}\mb{k}}(-\mathcal{C}_{xy}\mb{r}) \\
    i\psi^{b*}_{\mathcal{C}_{xy}\mb{k}}(-\mathcal{C}_{xy}\mb{r}) \\
    -i\chi^{b*}_{\mathcal{C}_{xy}\mb{k}}(-\mathcal{C}_{xy}\mb{r})
    \end{pmatrix}.
\end{equation}
If the momentum $\mb{k}$ is on the rotation axis, then $\mathcal{C}_{xy}\mb{k}=\mb{k}$, the above equation becomes
\begin{equation}
    \begin{pmatrix}
    \psi^t_{\mb{k}}(\mb{r}) \\
    \chi^t_{\mb{k}}(\mb{r}) \\
    \psi^b_{\mb{k}}(\mb{r}) \\
    \chi^b_{\mb{k}}(\mb{r})
    \end{pmatrix}
    =
    \begin{pmatrix}
    i\psi^{t*}_{\mb{k}}(-\mathcal{C}_{xy}\mb{r}) \\
    -i\chi^{t*}_{\mb{k}}(-\mathcal{C}_{xy}\mb{r}) \\
    i\psi^{b*}_{\mb{k}}(-\mathcal{C}_{xy}\mb{r}) \\
    -i\chi^{b*}_{\mb{k}}(-\mathcal{C}_{xy}\mb{r})
    \end{pmatrix}.
    \label{wave_on_axis}
\end{equation}
Since the position $\mb{r}$ and $\mathcal{C}_{xy}\mb{r}$ are symmetric to the line $y=-x$, Eq. \ref{wave_on_axis} tells us that the module of each component of the wavefunction is symmetric to the $y=-x$ line. If there is only one node of the wavefunction in real space, it must locates on the $y=-x$ line. So when we vary $\mb{k}$ on the $k_x=k_y$ line in momentum space, the node in real space can only move on the $y=-x$ line in real space, as we shown in the main text.

The analysis of the hexagonal lattice wavefunction is the same, just with the rotation axis and then the symmetry operator changed.

\section{wave function of two-fold degenerate flat band}

The theta functions $\theta_{a,b}(z|\tau)$ can be defined as
\begin{equation}
    \theta_{a,b}(z|\tau)=\sum_{n=-\infty}^{\infty}e^{i\pi (n+a)^2\tau+i2\pi(n+a)(z+b)}
\end{equation}
where $z,\tau\in\mathbb{Z}$ and $a,b\in\mathbb{R}$. The above theta function has the properties:
\begin{equation}
\begin{aligned}
    &\text{Property I:}& \quad &\theta_{a,b}(z+1|\tau)=e^{i2\pi a}\theta_{a,b}(z|\tau) \\
    &\text{Property II:}& \quad &\theta_{a,b}(z+\tau|\tau)=e^{-i\pi \tau-i2\pi(z+b)}\theta_{a,b}(z|\tau) \\
    &\text{Property III:}& \quad &\theta_{a,b}(z_0|\tau)=0,\quad z_0=(\frac{1}{2}+m-a)\tau+(\frac{1}{2}+n-b)
\end{aligned}
\label{theta_properties}
\end{equation}
where $m,n\in\mathbb{Z}$ are arbitrary integers. The first two properites give the quasi-periodicity of the theta functions, and the third one gives the zeros.

We construct the wave function as
$$
\psi_\mb{k}(\mb{r})=f_{\mb{k}}(z)\psi_{\mb{H}}(\mb{r})
$$
where $\mb{H}$ is the high symmetry point and $z=x+iy$. The wave function must satisify the Bloch condition
\begin{equation}
\begin{aligned}
    & \psi_{\mb{k}}(\mb{r}+\mb{L})=e^{i\mb{k}\cdot\mb{L}}\psi_{\mb{k}}(\mb{r}) \\
    \Longrightarrow &
    f_\mb{k}(\mb{r}+\mb{L})\psi_\mb{H}(\mb{r}+\mb{L})=e^{i\mb{k}\cdot\mb{L}}f_\mb{k}(\mb{r})\psi_{\mb{H}}(\mb{r}) \\
    \Longrightarrow &
    f_\mb{k}(\mb{r}+\mb{L})e^{i\mb{H}\cdot\mb{L}}\psi_\mb{H}(\mb{r})=e^{i\mb{k}\cdot\mb{L}}f_\mb{k}(\mb{r})\psi_{\mb{H}}(\mb{r}) \\
    \Longrightarrow &
    f_\mb{k}(\mb{r}+\mb{L})=e^{i(\mb{k}-\mb{H})\cdot\mb{L}}f_\mb{k}(\mb{r})
\end{aligned}
\end{equation}
This is the Bloch boundary condition for $f_\mb{k}(z)$. We can construct $f_\mb{k}(z)$ as
\begin{equation}
f_\mb{k}(z)=\frac{\theta_{a,b}(\frac{z}{L_1}|\frac{L_2}{L_1})}{\theta_{e,f}(\frac{z}{L_1}|\frac{L_2}{L_1})}
\end{equation}
where $z=x+iy$ and $L_1=L_{1x}+iL_{1y}$, $L_2=L_{2x}+iL_{2y}$. Since
\begin{equation}
\mb{L}=n\mb{L}_1+m\mb{L}_2, \Longrightarrow L=nL_1+mL_2
\end{equation}
using property I and property II in Eq. \ref{theta_properties} we can find
\begin{equation}
\begin{aligned}
    f_\mb{k}(z+L) =& \frac{\theta_{a,b}(\frac{z+L}{L_1}|\frac{L_2}{L_1})}{\theta_{e,f}(\frac{z+L}{L_1}|\frac{L_2}{L_1})} \\
    =& \frac{e^{n(i2\pi a)}e^{m(-i\pi\frac{L_2}{L_1}-i2\pi(\frac{z}{L_1}+b))}\theta_{a,b}(\frac{z}{L_1}|\frac{L_2}{L_1})} {e^{n(i2\pi e)}e^{m(-i\pi\frac{L_2}{L_1}-i2\pi(\frac{z}{L_1}+f))}\theta_{e,f}(\frac{z}{L_1}|\frac{L_2}{L_1})} \\
    =& e^{n(i2\pi(a-e))}e^{m(-i2\pi(b-f))}f_\mb{k}(z)
\end{aligned}
\end{equation}
so we have the condition
\begin{equation}
\begin{aligned}
    e^{n(i2\pi(a-e))}e^{m(-i2\pi(b-f))}=&e^{i(\mb{k}-\mb{H})\cdot\mb{L}} \\
    =& e^{i(\mb{k}-\mb{H})\cdot(n\mb{L}_1+m\mb{L}_2)} \\
    =& e^{n(i(\mb{k}-\mb{H})\cdot\mb{L}_1)}e^{m(i(\mb{k}-\mb{H})\cdot\mb{L}_2)}
\end{aligned}
\end{equation}
Finally we get the relation between the parameters
\begin{equation}
\begin{aligned}
    a-e=& \frac{(\mb{k}-\mb{H})\cdot\mb{L}_1}{2\pi} \\
    b-f=& -\frac{(\mb{k}-\mb{H})\cdot\mb{L}_2}{2\pi}
\end{aligned}
\end{equation}

Using the property III in Eq. \ref{theta_properties}, the zeros of theta function on the denominator are
\begin{equation}
    \theta_{e,f}(\frac{z_0}{L_1}|\frac{L_2}{L_1})=0, \quad
z_0=(\frac{1}{2}+m-e)L_2+(\frac{1}{2}+n-f)L_1
\end{equation}
these zeros must be canceled by the zero energy state $\psi_{\mb{H}}(\mb{r})$, or the wave function will blow up at this point. That is, $\psi_{\mb{H}}(\mb{r})$ must also have a zero at the position corresponding to $z_0$
\begin{equation}
    \psi_\mb{H}(\mb{r}_0)=0, \quad \mb{r}_0=(\frac{1}{2}+n-f)\mb{L}_1+(\frac{1}{2}+m-e)\mb{L}_2
\end{equation}
In practice, we first observe the zeros of $\psi_\mb{H}(\mb{r})$
\begin{equation}
\psi_\mb{H}(\mb{r}_0)=0, \quad \mb{r}_0=c_{\text{I}}\mb{L}_1+c_{\text{II}}\mb{L}_2
\end{equation}
then we fix the value of $e,f$ according to $\mb{r}_0$
\begin{align}
    \begin{cases}
    c_{\text{I}}=\frac{1}{2}+n-f \\
    c_{\text{II}}=\frac{1}{2}+m-e
    \end{cases}
    \Longrightarrow
    \begin{cases}
    f=\frac{1}{2}+n-c_{\text{I}} \\
    e=\frac{1}{2}+m-c_{\text{II}}
    \end{cases}
\end{align}
then we can fix $a,b$
\begin{equation}  
    \begin{cases}
    a-e=\frac{(\mb{k}-\mb{H})\cdot\mb{L}_1}{2\pi} \\
    b-f=-\frac{(\mb{k}-\mb{H})\cdot\mb{L}_2}{2\pi}
    \end{cases}
    \Longrightarrow
    \begin{cases}
    a=\frac{(\mb{k}-\mb{H})\cdot\mb{L}_1}{2\pi}+e \\
    b=-\frac{(\mb{k}-\mb{H})\cdot\mb{L}_2}{2\pi}+f
    \end{cases}
\end{equation}
the zeros of the numerator are
\begin{equation}
\begin{aligned}
    \theta_{a,b}(\frac{z_{\mb{k}}}{L_1}|\frac{L_2}{L_1})=0, \quad z_{\mb{k}}=(\frac{1}{2}+p-a)L_2+(\frac{1}{2}+q-b)L_1
\end{aligned}
\end{equation}
the corresponding location in real space is
\begin{equation}
    \psi_\mb{k}(\mb{r}_{\mb{k}})=\frac{\theta_{a,b}(\frac{z_{\mb{k}}}{L_1}|\frac{L_2}{L_1})}{\theta_{e,f}(\frac{z_{\mb{k}}}{L_1}|\frac{L_2}{L_1})}\psi_\mb{H}(\mb{r}_{\mb{k}})=0 \\
    \quad \mb{r}_{\mb{k}}=(\frac{1}{2}+q-b)\mb{L}_1+(\frac{1}{2}+p-a)\mb{L}_2
\end{equation}
since $a,b$ are functions of $\mb{k}$, we would observe a moving zero as we vary $\mb{k}$.

\section{Wave Function Of Four/Six-Fold Degenerate Flat Bands In Hexagonal Lattice}

In this section, we construct the wavefunction of the four-fold and six-fold degenerate flat bands from theta functions, and show that the construction coincide with that the numerical results.

\subsection{Wavefunction of four-fold degenerate flat band}

When checking the wavefunctions of the four-fold degenerate flat bands, we can observe two moving nodes, which implies that the wave function might be constructed by the product of two theta functions, so we can assume:
\begin{equation}
    \psi_{\mb{k}}(\mb{r})=\frac{\theta_{a,b}(\frac{z}{L_1}|\frac{L_2}{L_1})}{\theta_{e,f}(\frac{z}{L_1}|\frac{L_2}{L_1})} \frac{\theta_{a^\prime,b^\prime}(\frac{z}{L_1}|\frac{L_2}{L_1})}{\theta_{e^\prime,f^\prime}(\frac{z}{L_1}|\frac{L_2}{L_1})} \psi_{\mb{H}}(\mb{r})
    \label{2_theta_wavefunction}
\end{equation}
where we define $z=x+iy$ and $L_i=\mb{L}_{ix}+i\mb{L}_{iy}$. This construction satistifies the Bloch condition, and the zeros of the theta functions in the denominator must be canceled by the zeros of $\psi_{\mb{H}}(\mb{r})$. Suppose the nodes of the wavefunction are at
\begin{equation}
    \begin{cases}
    \psi_{\mb{\mb{H}}}(\mb{r}_0)=0, \quad \mb{r}_0=c_{\text{I}}\mb{L}_1+c_{\text{II}}\mb{L}_2 \\
    \psi_{\mb{\mb{H}}}(\mb{r}_0^\prime)=0, \quad \mb{r}^{\prime}_0=c_{\text{I}}^\prime\mb{L}_1+c_{\text{II}}^\prime\mb{L}_2
    \end{cases}
\end{equation}
the parameters $e,f$ and $e^\prime,f^\prime$ are decided by
\begin{equation}
    \begin{cases}
    f=\frac{1}{2}+n-c_{\text{I}} \\
    e=\frac{1}{2}+m-c_{\text{II}}
    \end{cases}
    \quad
    \begin{cases}
    f^\prime=\frac{1}{2}+n^\prime-c_\text{I}^\prime \\
    e^\prime=\frac{1}{2}+m^\prime-c_{\text{II}}^\prime
    \end{cases}
    \label{ef_4band}
\end{equation}
Here, $m,n,m^{\prime},n^{\prime} \in \mathbb{Z}$ are arbitrary integers. Choose an arbitrary $\mb{k}$ point in MBZ, we can calculate the wavefunction $\psi_{\mb{k}}(\mb{r})$ numerically and find its nodes. In the wavefunctions of the four-fold flat bands, we can find two zeros which moves with the variation of $\mb{k}$, that is
\begin{equation}
    \begin{cases}
    \psi_{\mb{k}}(\mb{r}_{\mb{k}})=0, \quad \mb{r}_{\mb{k}}=c_{\mb{k},\text{I}}\mb{L}_1+c_{\mb{k},\text{II}}\mb{L}_2 \\
    \psi_{\mb{k}}(\mb{r}_{\mb{k}}^\prime)=0, \quad \mb{r}^{\prime}_{\mb{k}}=c_{\mb{k},\text{I}}^\prime\mb{L}_1+c_{\mb{k},\text{II}}^\prime\mb{L}_2
    \end{cases}
\end{equation}
these two zeros must coincide with the that of the two theta functions on the numerator in Eq. \ref{2_theta_wavefunction}, so we can decide the parameters $a,b,a^{\prime},b^{\prime}$
\begin{equation}
    \begin{cases}
    b=\frac{1}{2}+q-c_{\mb{k},\text{I}} \\
    a=\frac{1}{2}+p-c_{\mb{k},\text{II}}
    \end{cases}
    \quad
    \begin{cases}
    b^\prime=\frac{1}{2}+q^\prime-c_{\mb{k},\text{I}}^\prime \\
    a^\prime=\frac{1}{2}+p^\prime-c_{\mb{k},\text{II}}^\prime
    \end{cases}
    \label{ab_4band}
\end{equation}
Again, $p,q,p^{\prime},q^{\prime}\in \mathbb{Z}$ are arbitrary integers.

To get the numerical wavefunction, we can calculate the momentum space wavefunction by diagonalizing the momentum space Hamiltonian $H(\mb{k})$, and then take Fourier transform to the real space. To get the wavefunction from the theta function construction, we first need to calculate the wavefunction $\Psi_{\mb{H}}(\mb{r})$ numerically, at the same time we can find its zeros to decide the parameters $e,f,e^{\prime},f^{\prime}$ using Eq. \ref{ef_4band}. Then we choose an arbitrary $\mb{k}$ to calculate the wavefunction $\Psi_{\mb{k}}(\mb{r})$ numerically to find its nodes, and we can get the parameters $a, b, a^{\prime},b^{\prime}$ using Eq. \ref{ab_4band}. Then, we can use these parameters to get the wavefunction $\Psi_{\mb{k}}(\mb{r})$ again from Eq. \ref{2_theta_wavefunction}, and compare with the numerical result. 

Here we show an example of the twisted bialyer hexagonal lattice with a quadratic node at the Gamma point. When only NN hopping is included, at the second magic value $\alpha^h_{1,2}=0.719$, the MFBs are four fold degenerate as shown in the main text, we show one of the wavefunctions of the four bands. We calculate $\Psi_{\mb{K}^{\prime}_{\text{MBZ}}}(\mb{r})$ numerically and find the two nodes, then we can get the parameters $e,f,e^{\prime},f^{\prime}$
\begin{equation}
    \begin{cases}
    \mb{r}_0=(-0.05125,0) \\
    \mb{r}_0^{\prime}=(-0.236713610367747;-0.5)
    \end{cases}
    \Longrightarrow
    \begin{cases}
    e=0.4704 \\
    f=0.4704 \\
    \end{cases}
    \begin{cases}
    e^{\prime}=-0.1367 \\
    f^{\prime}=0.8633
    \end{cases}
\end{equation}
Then we choose $\mb{k}=0.6\mb{K}^{\prime}_{\text{MBZ}}$, by numerical calculation we can find its zeros and the parameters $a,b,a^{\prime},b^{\prime}$ are
\begin{equation}
    \begin{cases}
    \mb{r}_{\mb{k}}=(-0.378750000000000,0) \\
    \mb{r}_{\mb{k}}^{\prime}=(-0.140007440278484;-0.5)
    \end{cases}
    \Longrightarrow
    \begin{cases}
    a=0.2813 \\
    b=0.2813 \\
    \end{cases}
    \begin{cases}
    a^{\prime}=-0.0808 \\
    b^{\prime}=0.9192
    \end{cases}
\end{equation}
We show the module of the wave function along a high symmetry path in Fig. \ref{SM_fig4}, the theta function construction concides with the numerical results perfectly.

\begin{figure}[H]
    \centering
    \includegraphics[scale=0.5]{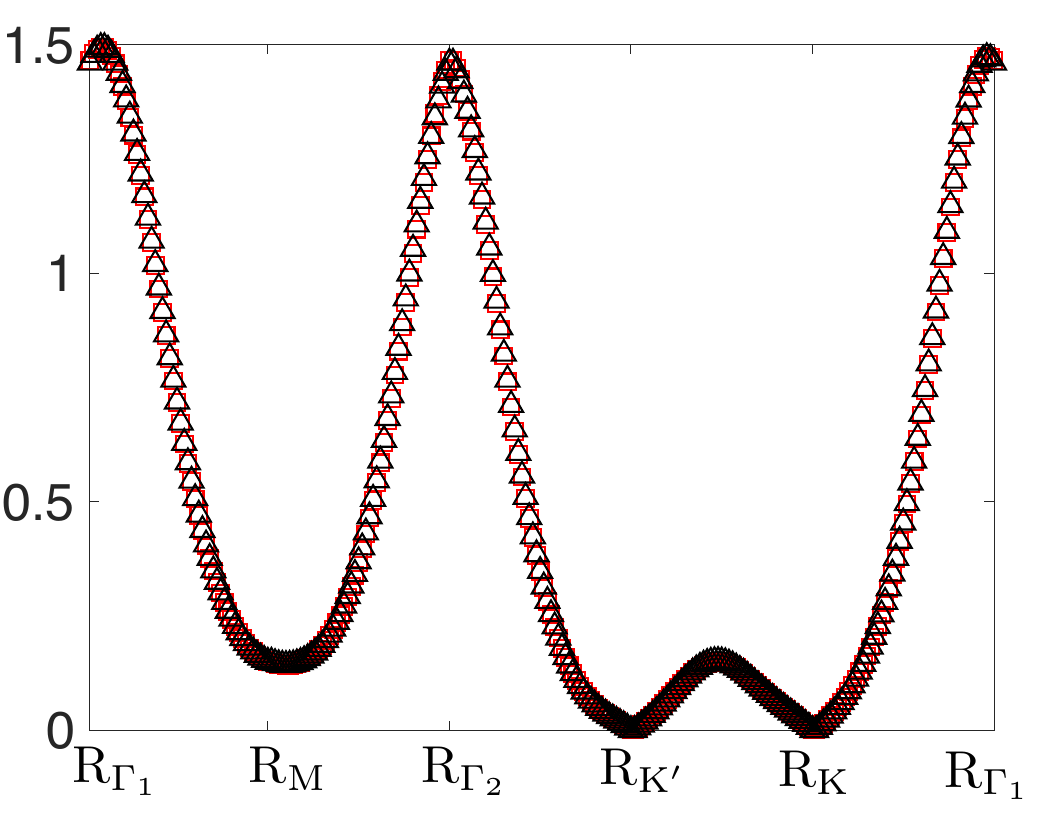}
    \caption{The module of the wavefunction $\Psi_{\mb{k}}(\mb{r})$ from the numerical calculation and the theta function construction. The red square are data from the numerical results, the black triangle are the data from theta function construction Eq. \ref{2_theta_wavefunction}.}
    \label{SM_fig4}
\end{figure}

\subsection{Wavefunction of six-fold degenerate flat band}

In sixfold degenerate flat band we can observe three moving nodes, so we can construct the wavefunction with three theta functions
\begin{equation}
    \psi_{\mb{k}}(\mb{r})=
    \frac{\theta_{a,b}(\frac{z}{L_1}|\frac{L_2}{L_1})}{\theta_{e,f}(\frac{z}{L_1}|\frac{L_2}{L_1})} 
    \frac{\theta_{a^{\prime},b^{\prime}}(\frac{z}{L_1}|\frac{L_2}{L_1})}{\theta_{e^{\prime},f^{\prime}}(\frac{z}{L_1}|\frac{L_2}{L_1})} 
    \frac{\theta_{a^{\dprime},b^{\dprime}}(\frac{z}{L_1}|\frac{L_2}{L_1})}{\theta_{e^{\dprime},f^{\dprime}}(\frac{z}{L_1}|\frac{L_2}{L_1})} 
    \psi_{\mb{H}}(\mb{r})
    \label{3_theta_wavefunction}
\end{equation}
where we define $z=x+iy$ and $L_i=\mb{L}_{ix}+i\mb{L}_{iy}$. The zeros of the theta functions in the denominator must be canceled by the zeros of $\psi_{\mb{H}}(\mb{r})$. Suppose the nodes of the wavefunction are at
\begin{equation}
    \begin{cases}
    \psi_{\mb{\mb{H}}}(\mb{r}_0)=0, \quad \mb{r}_0=c_{\text{I}}\mb{L}_1+c_{\text{II}}\mb{L}_2 \\
    \psi_{\mb{\mb{H}}}(\mb{r}_0^\prime)=0, \quad \mb{r}^{\prime}_0=c_{\text{I}}^\prime\mb{L}_1+c_{\text{II}}^\prime\mb{L}_2 \\
    \psi_{\mb{\mb{H}}}(\mb{r}_0^{\prime\prime})=0, \quad \mb{r}^{\prime\prime}_0=c_{\text{I}}^{\prime\prime}\mb{L}_1+c_{\text{II}}^{\prime\prime}\mb{L}_2
    \end{cases}
\end{equation}
the parameters $e,f$, $e^\prime,f^\prime$ and $e^{\prime\prime},f^{\prime\prime}$are decided by
\begin{equation}
    \begin{cases}
    f=\frac{1}{2}+n-c_{\text{I}} \\
    e=\frac{1}{2}+m-c_{\text{II}}
    \end{cases}
    \quad
    \begin{cases}
    f^\prime=\frac{1}{2}+n^\prime-c_\text{I}^\prime \\
    e^\prime=\frac{1}{2}+m^\prime-c_{\text{II}}^\prime
    \end{cases}
    \quad
    \begin{cases}
    f^{\prime\prime}=\frac{1}{2}+n^{\prime\prime}-c_\text{I}^{\prime\prime} \\
    e^{\prime\prime}=\frac{1}{2}+m^{\prime\prime}-c_{\text{II}}^{\prime\prime}
    \end{cases}
\end{equation}
Here, $m,n,m^{\prime},n^{\prime},m^{\prime\prime},n^{\prime\prime} \in \mathbb{Z}$ are arbitrary integers. Choose an arbitrary $\mb{k}$ point in MBZ, we can calculate the wavefunction $\Psi_{\mb{k}}(\mb{r})$ numerically and find its nodes. In the wavefunctions of the four-fold flat bands, we can find three zeros which moves with the variation of $\mb{k}$, that is
\begin{equation}
    \begin{cases}
    \psi_{\mb{k}}(\mb{r}_{\mb{k}})=0, \quad \mb{r}_{\mb{k}}=c_{\mb{k},\text{I}}\mb{L}_1+c_{\mb{k},\text{II}}\mb{L}_2 \\
    \psi_{\mb{k}}(\mb{r}_{\mb{k}}^\prime)=0, \quad \mb{r}^{\prime}_{\mb{k}}=c_{\mb{k},\text{I}}^\prime\mb{L}_1+c_{\mb{k},\text{II}}^\prime\mb{L}_2 \\
    \psi_{\mb{k}}(\mb{r}_{\mb{k}}^{\prime\prime})=0, \quad \mb{r}^{\prime\prime}_{\mb{k}}=c_{\mb{k},\text{I}}^{\prime\prime}\mb{L}_1+c_{\mb{k},\text{II}}^{\prime\prime}\mb{L}_2
    \end{cases}
\end{equation}
these two zeros must coincide with the that of the two theta functions on the numerator in Eq. \ref{3_theta_wavefunction}, so we can decide the parameters $a,b,a^{\prime},b^{\prime},a^{\prime\prime},b^{\prime\prime}$
\begin{equation}
    \begin{cases}
    b=\frac{1}{2}+q-c_{\mb{k},\text{I}} \\
    a=\frac{1}{2}+p-c_{\mb{k},\text{II}}
    \end{cases}
    \quad
    \begin{cases}
    b^\prime=\frac{1}{2}+q^\prime-c_{\mb{k},\text{I}}^\prime \\
    a^\prime=\frac{1}{2}+p^\prime-c_{\mb{k},\text{II}}^\prime
    \end{cases}
    \quad
    \begin{cases}
    b^{\prime\prime}=\frac{1}{2}+q^{\prime\prime}-c_{\mb{k},\text{I}}^{\prime\prime} \\
    a^{\prime\prime}=\frac{1}{2}+p^{\prime\prime}-c_{\mb{k},\text{II}}^{\prime\prime}
    \end{cases}
    \label{ab_6band}
\end{equation}
Again, $p,q,p^{\prime},q^{\prime},p^{\prime\prime},q^{\prime\prime}\in \mathbb{Z}$ are arbitrary integers. Then we can calculate the wavefunction from Eq. \ref{3_theta_wavefunction}, and compare this result with the numerical result to verify the construction is whether correct or not. 

Here we show an example of the twisted bialyer hexagonal lattice with a quadratic node at the Gamma point. When NN and NNN hopping are both included, at the third magic value $\alpha^{\prime h}_{1,3}=1.162$, the MFBs are six fold degenerate as shown in the main text, we show the one of wavefunctions of the six bands. We calculate $\Psi_{\mb{K}^{\prime}_{\text{MBZ}}}(\mb{r})$ numerically and find the three nodes, then we can get the parameters $e,f,e^{\prime},f^{\prime}, e^{\prime\prime},f^{\prime\prime}$ using Eq. \ref{ef_6band}
\begin{equation}
    \begin{cases}
    \mb{r}_0=(0.170318329410940,-0.5) \\
    \mb{r}_0^{\prime}=(-0.23,-0.205) \\
    \mb{r}_0^{\prime}=(-0.23,0.205)
    \end{cases}
    \Longrightarrow
    \begin{cases}
    e=0.0983 \\
    f=1.0983 \\
    \end{cases}
    \begin{cases}
    e^{\prime}=0.1622 \\
    f^{\prime}=0.5722
    \end{cases}
    \begin{cases}
    e^{\prime\prime}=0.5722 \\
    f^{\prime\prime}=0.1622
    \end{cases}
\end{equation}
Then we choose $\mb{k}=0.6\mb{K}^{\prime}_{\text{MBZ}}$, by numerical calculation we can find its zeros, and then we can get the parameters $a,b,a^{\prime},b^{\prime},a^{\prime\prime},b^{\prime\prime}$ using Eq. \ref{ab_6band}
\begin{equation}
    \begin{cases}
    \mb{r}_{\mb{k}}=(0.065673593120320;-0.5) \\
    \mb{r}_{\mb{k}}^{\prime}=(-0.235,-0.165) \\
    \mb{r}_{\mb{k}}^{\prime\prime}=(-0.235,0.165) 
    \end{cases}
    \Longrightarrow
    \begin{cases}
    a=0.0379 \\
    b=1.0379 \\
    \end{cases}
    \begin{cases}
    a^{\prime}=0.1993 \\
    b^{\prime}=0.5293
    \end{cases}
    \begin{cases}
    a^{\prime\prime}=0.5293 \\
    b^{\prime\prime}=0.1993
    \end{cases}
    \label{ef_6band}
\end{equation}
We show the module of the wave function along a high symmetry path in Fig. \ref{SM_fig5}, the theta function construction concides with the numerical results perfectly.

\begin{figure}[H]
    \centering
    \includegraphics[scale=0.5]{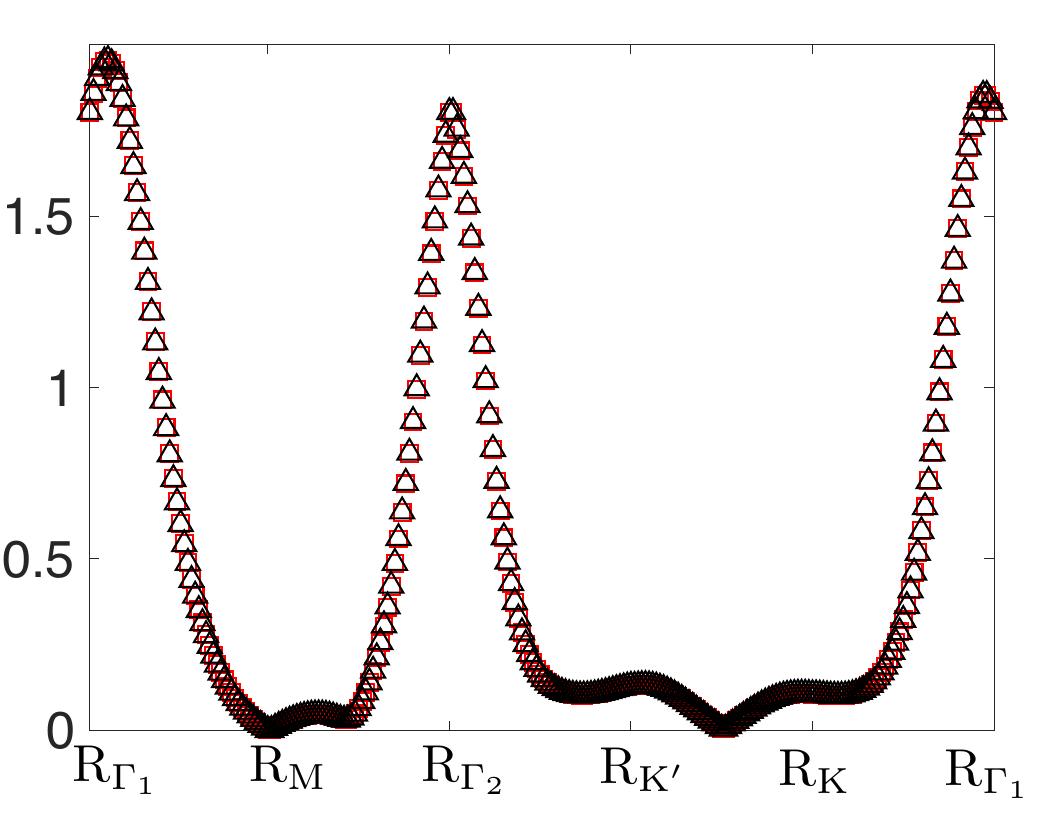}
    \caption{The module of the wavefunction $\Psi_{\mb{k}}(\mb{r})$ from the numerical calculation and the theta function construction. The red square are data from the numerical results, the black triangle are the data from theta function construction Eq. \ref{3_theta_wavefunction}.}
    \label{SM_fig5}
\end{figure}

\end{widetext}

\end{document}